\documentclass[12pt]{article}

\begin{document}
\newenvironment{Eqnarray}%
         {\arraycolsep 0.14em\begin{eqnarray}}{\end{eqnarray}}

\def\slash#1{{\rlap{$#1$} \thinspace/}}

\begin{flushright} 
March, 2001 \\
KEK-TH-752 \\
TIT/HEP-465 \\
\end{flushright} 

\hspace{1cm}

\begin{Large}
       \vspace{1cm}
  \begin{center}
   {Noncommutative Gauge Theories on\\
Fuzzy Sphere and Fuzzy Torus from Matrix Model}      \\
  \end{center}
    \end{Large}

  \vspace{0.8cm}

\begin{center}
YUSUKE KIMURA
\footnote{e-mail address : kimuray@post.kek.jp }

\vspace{0.8cm} 
{\it  High Energy Accelerator Research Organization (KEK),}\\
{\it Tsukuba, Ibaraki 305-0801, Japan} \\  
and \\
{\it  Department of Physics, Tokyo Institute of Technology, \\
Oh-okayama, Meguro-ku, Tokyo 152, Japan}\\
\end{center}
\vspace{0.6cm}

\begin{abstract}
\noindent
\end{abstract}
\hspace{0.2cm}
We consider a reduced model of four-dimensional 
Yang-Mills theory with a mass term. 
This matrix model has two classical solutions, 
two-dimensional fuzzy sphere 
and two-dimensional fuzzy torus. 
These classical solutions are constructed 
by embedding them into 
three or four dimensional flat space. 
They exist for finite size matrices, 
that is, 
the number of the quantum on these manifolds is finite. 
Noncommutative gauge theories on 
these noncommutative manifolds
are derived by expanding 
the model around these classical solutions 
and studied by 
taking two large $N$ limits, 
a commutative limit and a large radius limit. 
The behaviors of gauge invariant
operators are also discussed. 

\newpage


\section{Introduction}
\hspace{0.4cm}


After the discovery of the D-brane, our understanding 
of string theory has changed drastically. 
The development of D-brane physics\cite{pol,Taylor} 
leads to several proposals 
to formulate nonperturbative aspects 
of string theory. 
One of the notable properties of D-branes is that 
a system of $N$ coincident D-branes has collective coordinates 
which are described by $N \times N $ matrices 
and low energy dynamics of D-branes
are described by $U(N)$ supersymmetric gauge theories. 
This idea leads to 
several kinds of matrix models 
which have been proposed 
for the constructive definition 
of  string theory or M-theory\cite{BFSS,IKKT}. 
IIB matrix model\cite{IKKT} is one of these proposals.
It is  a large $N$ reduced model\cite{reduced} 
of ten-dimensional supersymmetric 
Yang-Mills theory and 
the action has a matrix regularized form of 
the Green-Schwarz action of IIB superstring. 

In the matrix model, 
eigenvalues of bosonic variables are interpreted 
as spacetime coordinates and  
matter and even spacetime are dynamically emerged out 
of matrices\cite{AIKKT,AIKKTT,AIIKKT}. 
Therefore noncommutative geometry appears naturally
from the matrix model. 
The idea of the noncommutative geometry is to modify the microscopic 
structure of the spacetime. 
This modification is implemented by replacing fields 
on the spacetime by matrices. 
Yang-Mills theories in noncommutative space 
is first appeared within the framework of 
troidai compactification of the matrix model\cite{CDS}. 
In string theory, 
it is pointed out that 
the world volume theory on D-branes 
with a constant NS-NS two-form background 
is described by noncommutative Yang-Mills theory\cite{SW}.
It was shown\cite{Li,AIIKKT,IIKK} that 
noncommutative Yang-Mills theories in flat backgrounds  
are obtained by expanding the matrix model 
around a flat noncommutative background. 
Fields on the background are appeared as 
the fluctuation around the background from the matrices. 
This implies  
the unification of spacetime and fields. 
Lattice version of noncommutative gauge theories is 
formulated in \cite{BM,AMNS}.


The IIB matrix model has only flat noncommutative 
backgrounds as classical solutions. 
To describe a curved spacetime 
and a compact spacetime 
is one of the important problems in matrix models. 
In our previous paper \cite{IKTW}, 
we treat a two-dimensional fuzzy sphere 
in a matrix model 
and show that expanding the model around 
the fuzzy sphere solution leads to a 
supersymmetric noncommutative 
gauge theory on the fuzzy sphere. 
Although ordinary matrix model does not have a fuzzy sphere 
as a classical solution,  
adding a Chern-Simons term to Yang-Mills reduced model
enable us to describe a fuzzy sphere as a classical 
solution and a noncommutative gauge theory 
on it as discussed in \cite{IKTW}. 
Owing to the extra term, the matrix model 
can describe a curved spacetime. 
In \cite{myers}, 
$N$ D-branes action in nontrivial backgrounds 
is considered and Chern-Simons term 
appeared from D$0$-branes action 
in a constant RR three-form potential. 
Owing to the Chern-Simons term, 
D$0$-branes expand into a noncommutative 
fuzzy sphere configuration. 
However, 
it is not well understood 
how the extra term is generated from 
the IIB matrix model. 

In this paper, we treat a fuzzy sphere and a fuzzy torus 
in a matrix model by the same manner 
as in \cite{Li,AIIKKT,IKTW}. 
We start with a bosonic matrix model which is given by 
\begin{equation}
 S= -\frac{1}{g^{2}} 
Tr \left( \frac{1}{4} \left[ A_{\mu} ,A_{\nu}\right] 
         \left[A^{\mu} ,A^{\nu} \right]  
     +\lambda^{2} A_{\mu}A^{\mu} \right) . \label{action}
\end{equation}
$A_{\mu}$ are $N \times N$ hermitian matrices. 
$\lambda$ is a dimensionful parameter which depends on $N$.
This is a reduced model, 
which is obtained by reducing the spacetime volume 
to a single point\cite{reduced}, 
of a four-dimensional bosonic Yang-Mills theory 
with a mass term\footnote{
We now comment on the mass term. 
This mass term corresponds to 
a negative mass. 
Therefore we cannot avoid an unstable mode which is originated from 
the negative mass term.}. 
This model possesses $SO(4)$ symmetry 
and gauge symmetry expressed 
by the following unitary transformation, 
\begin{equation}
A_{\mu} \rightarrow UA_{\mu}U^{\dagger}. \label{gaugetr}
\end{equation}
This model does not have translational symmetry of $A_{\mu}$ 
because of the mass term. 
In spite of this shortcoming, this model 
has an interesting property. 
The equation of motion 
of the action (\ref{action}) 
is given by 
\begin{equation}
[A_{\nu},[A_{\mu},A_{\nu}]] +2\lambda^{2} A_{\mu}=0. 
\label{classicalsolution}
\end{equation}
This equation of motion
\footnote{This type of the equation of motion 
is considered in \cite{Hoppe9702169} 
in the context of the BFSS matrix model.
To find solutions of this equation is 
related to the problem of embedding a compact 
two-dimensional surface into 
a three-dimensional sphere\cite{HoppeNicolai}. 
} 
has two classical solutions 
with different topology. 
One is a two-dimensional fuzzy sphere and 
the other is a two-dimensional fuzzy torus 
(These solutions are explained in section 2 and 
section 3 respectively). 
Both solutions are constructed by embedding them 
into three or four-dimensional  
flat spacetime. 
Flat backgrounds do not exist for finite $N$, 
while sphere and torus
\footnote{
Another construction of torus is 
to impose a periodic boundary condition
in a flat background\cite{Taylor, Taylor9611042}. 
This construction make us 
introduce infinite copies of the 
original matrices. 
Therefore we cannot construct for finite $N$.} 
backgrounds 
can exist 
for finite $N$. 
In the matrix model picture, $N$ represents the 
number of the quantum on the backgrounds (
or the number of D-instantons or D-particles).
As a flat background with infinite extent 
has the infinite number of the quantum on it, 
it does not exist for finite $N$. 
On the other hand, since 
the area of compact backgrounds 
is finite, the solution of compact backgrounds 
can be constructed for finite $N$.  

The goal of this paper is 
to consider a curved spacetime or a compact spacetime 
(a noncommutative sphere and a noncommutative torus)
\footnote{
Similar analysis in a tube configuration is shown 
in \cite{bak}.} 
in matrix models. 
We show that expanding the model around classical solutions 
leads to noncommutative gauge theories on the solutions.


This paper is organized as follows. 
Section 2 and 3 are devoted to 
the analysis of the noncommutative gauge theories 
on the fuzzy sphere and the fuzzy torus respectively. 
Section 2 is based on our previous paper \cite{IKTW}.
The relation between the model considered in 
the previous paper \cite{IKTW}
and the model used in this paper 
is also discussed. 
Construction of a noncommutative torus 
in terms of unitary matrices 
is well known. 
As the eigenvalue of unitary matrices are distributed 
over $S^{1}$, compactness is naturally described 
in terms of unitary matrices. 
In section 3, we construct a fuzzy torus
in terms of hermitian matrices 
by dividing a unitary matrix 
into two hermitian matrices. 
To investigate the noncommutative gauge theories
on these two noncommutative manifolds, 
we consider two large $N$ limits. 
One is a commutative limit and another is 
a large radius limit. 
The first limit gives ordinary gauge 
theories on a commutative sphere and 
a commutative torus. 
On the other hand, 
the second limit gives a noncommutative gauge 
theory on a noncommutative flat space. 
Although these two gauge theories are same 
from the matrix model point of view, 
two large $N$ limits distinguish 
these gauge theories. 
We observe the difference of the symmetry 
($SO(3)$ versus $SO(2)\times SO(2)$) 
by taking a commutative limit. 
In section4, 
the behavior of gauge invariant 
operators on a noncommutative sphere and  
a noncommutative torus is investigated.  
Section 5 is devoted to summary and discussions.


\section{Noncommutative gauge theory on noncommutative sphere}
\hspace{0.4cm}
In this section, we treat a noncommutative sphere. 
The fuzzy sphere
\cite{madore,GKP,WW,Klimcik,GM,GP,bala,KabatTay,Rey,Hoppe} 
can be constructed 
by introducing a cut off parameter $N-1$ 
for angular momentum of the spherical harmonics. 
The number of independent functions is 
$\sum_{l=0}^{N-1}(2l+1)=N^{2}$. 
Therefore, 
we can replace the functions by $N \times N$ hermitian matrices 
on the fuzzy sphere. 
Thus, the algebra on the fuzzy sphere becomes noncommutative. 

A noncommutative gauge theory on a noncommutative 
sphere is considered in our previous paper\cite{IKTW}. 
Chern-Simons term is added in this case. 
Since the construction of the noncommutative gauge theory  
is parallel between Chern-Simons term case 
and mass term case, 
we mainly follow our previous paper\cite{IKTW}. 
A noncommutative sphere is constructed 
by embedding it into ${\bf R}^{3}$. 
It  is represented by 
the following algebra, 

\begin{Eqnarray}
&&[\hat{x}_{i},\hat{x}_{j}]=i\alpha\epsilon_{ijk}\hat{x}_{k} 
\hspace{0.2cm} (i,j,k=1,2,3),   \cr 
&&\hat{x}_{4}=0. 
\label{clsol}
\end{Eqnarray}
$A_{\mu}=\hat{x}_{\mu}$ satisfy (\ref{classicalsolution}) 
if we set 
$\lambda^{2}=\alpha^{2}$. 
\noindent 
Indices $i,j,k$ are used for $1,2,3$ and 
$\mu, \nu, \tau$ 
for $1,2,3,4$. 
We impose the following condition for $\hat{x}_{i}$, 
\begin{equation}
\hat{x}_{1}\hat{x}_{1}+\hat{x}_{2}\hat{x}_{2}+\hat{x}_{3}\hat{x}_{3} =\rho^{2}.
\end{equation}
This solution preserves $SO(3)$ symmetry. 
In the $\alpha \rightarrow 0$ limit, $\hat{x}_{i}$ becomes commutative 
coordinates $x_{i}$: 
\begin{Eqnarray}
x_{1}&=&\rho \sin \theta \cos \phi \cr
x_{2}&=&\rho \sin \theta \sin \phi \cr
x_{3}&=&\rho \cos \theta ,
\end{Eqnarray}
\noindent 
where $\rho$ denotes the radius of the sphere. 
The metric tensor of the commutative sphere is 
\begin{Eqnarray}
ds^{2}&=&\rho^{2}d\theta^{2}
+\rho^{2}\sin^{2}\theta d\phi^{2} \cr 
&\equiv& \rho^{2} g_{ab}d\sigma^{a}d\sigma^{b}. 
\end{Eqnarray} 
Matrices $\hat{x_{i}}$ can be constructed from the 
generators of the $N$-dimensional irreducible representation 
of $SU(2)$ as 
\begin{equation}
\hat{x}_{i}=\alpha \hat{L}_{i} 
\label{coordinatemomentumsphere}
\end{equation}
with 
\begin{equation}
[\hat{L}_{i},\hat{L}_{j}]=i\epsilon_{ijk}\hat{L}_{k}. 
\label{su(2)}
\end{equation}
The radius of the sphere is given by the 
quadratic Casimir of $SU(2)$ as 
\begin{equation}
\rho^{2}=\frac{N^{2}-1}{4}\alpha^{2}. 
\end{equation}
The Plank constant on the fuzzy sphere, which 
represents the area occupied by the unit quantum 
on the fuzzy sphere, is given by 
\begin{equation}
\frac{4\pi\rho^{2}}{N}
=\frac{N^{2}-1}{N}\pi\alpha^{2}.  \label{plank}
\end{equation}
 
\vspace{0.5cm}

Now we show that an expansion of the model 
around the classical background 
(\ref{clsol}) by the similar procedure as in \cite{Li,AIIKKT} 
leads to a noncommutative Yang-Mills on a fuzzy sphere.
We first consider $U(1)$ noncommutative gauge theory on the fuzzy sphere. 
We expand the bosonic matrices around 
the classical solution (\ref{clsol}) as 
\begin{Eqnarray}
&&A_{i}= \hat{x}_{i}+\alpha\rho\hat{a}_{i}=
\alpha \rho(\frac{\hat{L}_{i}}{{\rho}}+\hat{a}_{i})
\hspace{0.2cm} \cr
&&A_{4}=\alpha \rho \hat{\phi}
\end{Eqnarray}
$\hat{a}_{i}$ and $\hat{\phi}$ are fields 
which propagate on the fuzzy sphere. 
A notable point is that the background $\hat{x}_{i}$ 
and the fields $\hat{a}_{i}$ and $\hat{\phi}$ 
are dynamically generated 
from matrix $A_{i}$ and 
they are treated 
on the same footing. 

We first study a correspondence 
between matrices and functions on a sphere. 
Ordinary functions on the sphere can be expanded by the spherical harmonics, 
\begin{Eqnarray}
a(\Omega)=\sum_{l=0}^{\infty}\sum_{m=-l}^{l}
a_{lm}Y_{lm}(\Omega),  \label{function}
\end{Eqnarray}
\noindent 
where 
\begin{equation}
Y_{lm}= \rho^{-l}\sum_{a}f_{a_{1},a_{2},\cdots a_{l}}^{(lm)}
x^{a_{1}}\cdots x^{a_{l}} \label{harmonics}
\end{equation}
\noindent 
is a spherical harmonics and 
$f_{a_{1},a_{2},\cdots a_{l}}$ is a traceless and 
symmetric tensor. 
The traceless condition comes from $x_{i}x_{i}=\rho^{2}$. 
The normalization of the spherical harmonics is fixed by 
\begin{equation}
\int \frac{d\Omega}{4\pi} Y^{\ast}_{l^{\prime}m^{\prime}}Y_{lm}
=\frac{1}{4\pi} \int_{0}^{2\pi}d\varphi \int_{0}^{\pi} \sin \theta d\theta 
Y^{\ast}_{l^{\prime}m^{\prime}}Y_{lm}
=\delta_{l^{\prime}l}\delta_{m^{\prime}m} . \label{nor1}
\end{equation}
\noindent 
Matrices on the fuzzy sphere, on the other hand, 
can be expanded 
by {\sl the noncommutative spherical harmonics} $\hat{Y}_{lm}$ 
as 
\begin{Eqnarray}
\hat{a}=\sum_{l=0}^{N-1}\sum_{m=-l}^{l}
a_{lm}\hat{Y}_{lm}.  \label{matrix}
\end{Eqnarray}
\noindent 
$\hat{Y}_{lm}$ is a $N\times N$ matrix and defined by 
\begin{equation}
\hat{Y}_{lm}=\rho^{-l}\sum_{a}f_{a_{1},a_{2},\cdots,a_{l}}^{lm}
\hat{x}^{a_{1}}\cdots\hat{x}^{a_{l}} , \label{nonspheri}
\end{equation}
\noindent 
where the same coefficient as (\ref{harmonics}) is used. 
Angular momentum $l$ is bounded at $l=N-1$ 
and these $\hat{Y}_{lm}$'s form 
a complete basis of $N\times N$ hermitian matrices.
From the symmetry of the indices, the ordering of $\hat{x}$ 
corresponds to 
the Weyl type ordering
\footnote{
In \cite{IKTW}, 
normal ordered basis is investigated 
by a stereographic projection 
from a sphere to a complex plane 
and Berezin type star product is 
obtained.}. 
A hermiticity condition requires that 
$a_{lm}^{\ast}=a_{l-m}$.
Normalization of the noncommutative spherical harmonics is given by  
\begin{equation}
\frac{1}{N}Tr(\hat{Y}^{\dagger}_{l^{\prime}m^{\prime}}
\hat{Y}_{lm})=\delta_{l^{\prime}l}\delta_{m^{\prime}m}. \label{normali}
\end{equation}
Let us consider a map from matrices to functions:   
\begin{equation}
\hat{a}=\sum_{l=0}^{N-1}\sum_{m=-l}^{l}
a_{lm}\hat{Y}_{lm}
 \rightarrow 
a(\Omega)=\sum_{l=0}^{N-1}\sum_{m=-l}^{l}
a_{lm}Y_{lm}(\Omega).  
\end{equation}
This map is formally given as 
\begin{equation}
a(\Omega)=\frac{1}{N}\sum_{lm}
Tr(\hat{Y}_{lm}^{\dagger}\hat{a})Y_{lm}(\Omega), 
\label{mapsphere}
\end{equation}

\noindent 
and correspondingly 
a product of matrices is mapped to {\it the star product} on the 
fuzzy sphere: 

\begin{Eqnarray}
a\star b(\Omega)=\frac{1}{N}\sum_{lm} 
Tr(\hat{Y}_{lm}^{\dagger}\hat{a}\hat{b})Y_{lm}(\Omega).   
\label{productmap}
\end{Eqnarray}
This product is noncommutative corresponding to the  
noncommutativity of matrices. 
The explicit form of this product is calculated 
in \cite{IKTW,Hoppe}. 
Let us consider the product of the two spherical harmonics, 
$\hat{Y}_{lm}$ and $\hat{Y}_{l^{\prime}m^{\prime}}$. 
We have required that the maximal 
value of $l$ is $N-1$. 
This product is expanded by the spherical harmonics and it contains 
$\hat{Y}_{l+l^{\prime}}$. 
We assume that $N$ is large such that 
$l+l^{\prime}$ does not exceed $N-1$.  
This assumption guarantees that 
the map (\ref{mapsphere}) is one to one. 

We next study derivative operators 
corresponding to the adjoint action of $\hat{L}_{i}$. 
An action of $Ad(\hat{L}_{3})$ is calculated as 
\begin{equation} 
Ad(\hat{L}_{3})\hat{a}=\sum_{lm}a_{lm}[\hat{L}_{3},\hat{Y} _{lm} ] 
=\sum_{lm}a_{lm}m\hat{Y} _{lm}.
\end{equation}
\noindent 
This property and $SO(3)$ symmetry gives 
the following correspondence: 
\begin{Eqnarray}
Ad(\hat{L}_{i}) \rightarrow 
L_{i} \equiv \frac{1}{i}\epsilon_{ijk}x_{j}\partial_{k}. \label{angope}
\end{Eqnarray}
\noindent 
The Laplacian on the fuzzy sphere is given by 
\begin{equation}
\frac{1}{\rho^{2}}Ad(\hat{L})^{2}\hat{a}
=\frac{1}{\rho^{2}}
\sum_{lm}a_{lm}[\hat{L}_{i}, [\hat{L}_{i},\hat{Y} _{lm} ]] 
=\sum_{lm}\frac{l(l+1)}{\rho^{2}}a_{lm}\hat{Y} _{lm}. 
\end{equation}
\noindent
We can rewrite $L_{i}$ in terms of Killing vectors 
on the sphere as 
\begin{Eqnarray}
L_{i}&=&-iK_{i}^{a}\partial_{a}.  
\end{Eqnarray}
The metric tensor is also given 
in terms of Killing vectors as 
\begin{Eqnarray}
g^{ab}=K_{i}^{a}K_{i}^{b}. 
\end{Eqnarray} 
The explicit forms of the Killing vectors are shown 
in the appendix. 

$Tr$ over matrices can be mapped to the integration over functions as 
\begin{equation}
\frac{1}{N}Tr (\hat{a})\rightarrow 
\int \frac{d\Omega}{4\pi}a(\Omega). 
\end{equation}

Let us expand the action (\ref{action}) around 
the classical solution (\ref{clsol}) and 
apply these mapping rules. 
The action becomes 

\begin{Eqnarray}
S&=&-\frac{1}{g^{2}} Tr(\frac{\alpha^{4}\rho^{4}}{4}
\hat{F}_{ij}\hat{F}_{ij}
+\frac{i\alpha^{3}\rho^{2}}{2}\epsilon_{ijk}
\hat{F}_{ij}A_{k}
-\frac{\alpha^{2}}{2}A_{i}A_{i} \cr 
&& \hspace{1cm} +\frac{\alpha^{4}\rho^{4}}{2}
[\frac{\hat{L}_{i}}{{\rho}}+\hat{a}_{i},\hat{\phi}]
[\frac{\hat{L}_{i}}{{\rho}}+\hat{a}_{i},\hat{\phi}])   
-\frac{\alpha^{2}}{g^{2}}TrA_{\mu}A_{\mu} \cr 
&\rightarrow&-\frac{\rho^{2}}{4g_{YM}^{2}}\int d\Omega(
F_{ij}F_{ij} +
2[\frac{L_{i}}{{\rho}}+a_{i},\phi]
[\frac{L_{i}}{{\rho}}+a_{i},\phi]
+\frac{4}{\rho^{2}}\phi^{2} 
)_{\star} \cr 
&&-\frac{3i}{2g_{YM}^{2}}
\epsilon_{ijk}\int d\Omega(
(L_{i}a_{j})a_{k}+\frac{\rho}{3}[a_{i},a_{j}]a_{k}
-\frac{i}{2}\epsilon_{ijl}a_{l}a_{k})_{\star} \cr
&&
-\frac{\pi}{g_{YM}^{2}}\frac{N^{2}}{2\rho^{2}}
\end{Eqnarray}
where $\hat{F}_{ij}$ is the field strength on the sphere 
and given by 
\begin{Eqnarray}
\hat{F}_{ij}&=&\frac{1}{\alpha^{2}\rho^{2}}
([A_{i},A_{j}]-i\alpha\epsilon_{ijk}A_{k}) \cr 
&=&[\frac{\hat{L}_{i}}{\rho},\hat{a}_{j}] 
-[\frac{\hat{L}_{j}}{\rho},\hat{a}_{i}]
+[\hat{a}_{i},\hat{a}_{j}]-\frac{1}{\rho}i\epsilon_{ijk}\hat{a}_{k} 
\end{Eqnarray} 
and this is mapped to the following function 
\begin{Eqnarray}
F_{ij}(\Omega)=\frac{1}{\rho}L_{i}a_{j}(\Omega)
 -\frac{1}{\rho}L_{j}a_{i}(\Omega)
+[a_{i}(\Omega),a_{j}(\Omega)]_{\star}
-\frac{1}{\rho}i\epsilon_{ijk}a_{k}(\Omega). 
\end{Eqnarray}
Gauge covariance of $F_{ij}$ is manifest 
from the viewpoint of the matrix model 
and $F_{ij}$ becomes zero when the fluctuations 
are set to zero. 
$(\hspace{0.4cm})_{\star}$ means that 
the products should be taken as the star product. 
The Yang-Mills coupling $g_{YM}^{2}$ 
is defined by 
\begin{equation}
g_{YM}^{2}=\frac{4\pi g^{2}}{N\alpha^{4}\rho^{2}}. 
\end{equation}
\noindent 
Thus we have obtained U(1) noncommutative gauge theory 
on the fuzzy sphere 
by expanding the matrix model around the fuzzy sphere solution 
and mapping the matrix model 
to the field theory. 

We have so far discussed the $U(1)$ noncommutative gauge theory 
on the fuzzy sphere. 
A generalization to $U(m)$ gauge group is realized 
by the following replacement: 
\begin{equation}
\hat{x}_{i} \rightarrow \hat{x}_{i}\otimes{\bf 1}_{m}. 
\end{equation}
\noindent 
$\hat{a}$ is also replaced as follows:
\begin{equation}
\hat{a} \rightarrow 
\sum_{a=1}^{m^{2}}\hat{a}^{a}\otimes T^{a}, 
\end{equation}
\noindent 
where $T^{a} (a=1,\cdots, m^{2})$ denote the generators of $U(m)$.
Then we obtain a $U(m)$ noncommutative gauge theory 
by the same procedure as the $U(1)$ case: 
\begin{Eqnarray}
S&=&-\frac{\rho^{2}}{4g_{YM}^{2}}tr\int d\Omega(
F_{ij}F_{ij} +
2[\frac{L_{i}}{{\rho}}+a_{i},\phi]
[\frac{L_{i}}{{\rho}}+a_{i},\phi]
+\frac{4}{\rho^{2}}\phi^{2} 
)_{\star} \cr 
&&-\frac{3i}{2g_{YM}^{2}}
\epsilon_{ijk}tr\int d\Omega(
(L_{i}a_{j})a_{k}+\frac{\rho}{3}[a_{i},a_{j}]a_{k}
-\frac{i}{2}\epsilon_{ijl}a_{l}a_{k})_{\star} \cr
&&-\frac{\pi}{g_{YM}^{2}}\frac{mN^{2}}{2\rho^{2}}
\end{Eqnarray}
\noindent 
where $tr$ is taken over $m \times m$ matrices. 

We next focus on the gauge symmetry of this action. 
The action (\ref{action}) is invariant under 
the unitary transformation (\ref{gaugetr}).  
Gauge symmetry of noncommutative gauge theories 
is included in the unitary transformation 
(\ref{gaugetr}) of the matrix model. 
For an infinitesimal transformation 
$U=\exp(i\hat{\lambda}) \sim 1+i\hat{\lambda}$ in (\ref{gaugetr}) 
where $\hat{\lambda}=\sum_{lm}\lambda_{lm}\hat{Y}_{lm}$, 
the fluctuation around the fixed background 
transforms as 
\begin{Eqnarray}
\hat{a}_{i} &\rightarrow& \hat{a}_{i} 
-\frac{i}{\rho}[\hat{L}_{i},\hat{\lambda}]
  +i[\hat{\lambda},\hat{a}_{i}]. \label{matrixgauge}  
\end{Eqnarray}
After mapping to functions, we have local gauge symmetry
\begin{Eqnarray}
a_{i}(\Omega) &\rightarrow& 
a_{i}(\Omega) -\frac{i}{\rho}L_{i}\lambda(\Omega)
  +i[\lambda(\Omega),a_{i}(\Omega)]_{\star}.
  \label{gatr}
\end{Eqnarray}

Let us discuss a scalar field 
which is a normal component of $\hat{a}_{i}$. 
Three fields $\hat{a}_{i}$ are defined in three dimensional 
space {\bf R$^{3}$} 
and contain a gauge field on $S^{2}$ 
and a scalar field as well. 
We define it by
\begin{Eqnarray}
\hat{\varphi}&\equiv&\frac{1}{2\alpha\rho}
(A_{i}A_{i}- \hat{x}_{i}\hat{x}_{i}) \cr 
&=&\frac{1}{2}
(\hat{x}_{i}\hat{a}_{i}+\hat{a}_{i}\hat{x}_{i} 
+\alpha\rho\hat{a}_{i}\hat{a}_{i}).  
\end{Eqnarray}
\noindent 
It transforms covariantly as an adjoint representation 
under the gauge transformation 
\begin{equation}
\hat{\varphi} \rightarrow 
\hat{\varphi} +i[\hat{\lambda}, \hat{\varphi}].
\end{equation}
Since the scalar field should become 
the radial component of $\hat{a}_{i}$ in the commutative limit, 
a naive choice is
$\hat{\varphi}_{0}=
(\hat{x}_{i}\hat{a}_{i}+\hat{a}_{i}\hat{x}_{i})/2$. 
For small fluctuations this field is 
the correct component of the field $\hat{a}_{i}$ but 
large fluctuations of $\hat{a}_{i}$ deform the shape 
of the sphere and $\hat{\varphi}_{0}$ can be no longer interpreted 
as the radial component of $\hat{a}_{i}$. 
This is a manifestation of the fact that matrix models or 
noncommutative gauge theories naturally unify spacetime and 
matter on the same footing.
An addition of the non-linear term 
$\hat{a}_{i}\hat{a}_{i}$ makes $\hat{\varphi}$ transform correctly 
as the scalar field in the adjoint representation.

\vspace{0.5cm}

We now consider a commutative limit. 
From (\ref{plank}), the commutative limit 
is realized by 
\begin{Eqnarray}
\rho=\mbox{fixed}, \hspace{0.4cm} 
g_{YM}=\mbox{fixed}, \hspace{0.4cm} 
N\rightarrow \infty. 
\end{Eqnarray}
In the commutative limit, 
the star product becomes the commutative product. 
The action becomes 
\begin{Eqnarray}
S=&&-\frac{\rho^{2}}{4g_{YM}^{2}}tr \int d\Omega(
F_{ij}F_{ij} +
2[\frac{L_{i}}{{\rho}}+a_{i},\phi]
[\frac{L_{i}}{{\rho}}+a_{i},\phi]
+\frac{4}{\rho^{2}}\phi^{2}) \cr 
&&-\frac{3i}{2g_{YM}^{2}}
\epsilon_{ijk} tr \int d\Omega (
(L_{i}a_{j})a_{k}+\frac{\rho}{3}[a_{i},a_{j}]a_{k}
-\frac{i}{2}\epsilon_{ijl}a_{l}a_{k}) \cr
&&
-\frac{\pi}{g_{YM}^{2}}\frac{N^{2}}{2\rho^{2}}. 
\label{commutativeactionsphere}
\end{Eqnarray}
Since the noncommutativity deforms 
the shape of the sphere as explained 
in the previous paragraph, 
the scalar field cannot be separated from 
the gauge field. 
In the commutative case, however, 
the scalar field $\varphi$ and the gauge field are separable 
from each other as in 
\begin{Eqnarray}
\rho a_{i}(\Omega)&=& 
= K_{i}^{a}b_{a}(\Omega) 
+ \frac{x_{i}}{\rho} \varphi(\Omega)  
\label{Killing}
\end{Eqnarray}
where $i=1,2,3$ and $a=\theta,\phi$. 
$b_{a}$ is a gauge field on the sphere.
The field strength is expressed in terms of 
the gauge field $b_{a}$ and the scalar field $\varphi$ as 
\begin{Eqnarray}
F_{ij}(\Omega) 
=\frac{1}{\rho^{2}}K_{i}^{a}K_{j}^{b}F_{ab} 
+\frac{i}{\rho^{2}}\epsilon_{ijk}x_{k}\varphi
+\frac{1}{\rho^{2}}x_{j}K_{i}^{a}D_{a}\varphi
-\frac{1}{\rho^{2}}x_{i}K_{j}^{a}D_{a}\varphi
\end{Eqnarray}
where 
$F_{ab}=\frac{1}{i}\partial_{a}b_{b}
-\frac{1}{i}\partial_{b}b_{a}+[b_{a},b_{b}]$ and 
$D_{a}=\frac{1}{i}\partial_{a}+[b_{a},\cdot]$. 
The action in 
(\ref{commutativeactionsphere}) 
is also rewritten as 
\begin{Eqnarray}
S=&&-\frac{1}{4g_{YM}^{2}\rho^{2}}
tr\int d\Omega(
K_{i}^{a}K_{j}^{b}K_{i}^{c}K_{j}^{d}
F_{ab}F_{cd}
+i2K_{i}^{a}K_{j}^{b}F_{ab}\epsilon_{ijk}
\frac{x_{k}}{\rho}\varphi \cr 
&&\hspace{1.5cm}
+2K_{i}^{a}K_{i}^{b}
(D_{a}\varphi)(D_{b}\varphi)
-2\varphi^{2} \cr 
&&\hspace{1.5cm}
+2\rho^{2}K_{i}^{a}K_{i}^{b}(D_{a}\phi)(D_{b}\phi)
+4\rho^{2}\phi^{2} \cr 
&&\hspace{1.5cm}
+4\rho^{2}K_{i}^{a}\frac{x_{i}}{\rho}(D_{a}\phi)([\varphi,\phi]) 
+2\rho^{2}\frac{x_{i}}{\rho}\frac{x_{i}}{\rho}([\varphi,\phi])^{2}
) \cr 
&&-\frac{3}{2g_{YM}^{2}\rho^{2}}
tr \int d\Omega (i
\epsilon_{ijk}K_{i}^{a}K_{j}^{b}
F_{ab}\frac{x_{k}}{\rho}\varphi
-\varphi^{2}) \cr
=&&-\frac{1}{4g_{YM}^{2}\rho^{2}}
tr\int d\Omega(
F_{ab}F^{ab}
+8i\frac{\epsilon^{ab}}{\sqrt{g}}F_{ab}\varphi \cr 
&&\hspace{1.0cm}
+2(D_{a}\varphi)(D^{a}\varphi)
-8\varphi^{2} 
+2\rho^{2}(D_{a}\phi)(D^{a}\phi)
+4\rho^{2}\phi^{2}
+2\rho^{2}([\varphi,\phi])^{2}
)  
\label{commutativeKillingaction} 
\end{Eqnarray}
\noindent 
where $\epsilon^{\theta\phi}=-\epsilon^{\phi\theta}=1$. 
The gauge transformation (\ref{gatr}) becomes 
\begin{Eqnarray}
b_{a} &\rightarrow& b_{a} - \partial_{a}\lambda  \cr 
\varphi &\rightarrow& \varphi  \cr 
\phi &\rightarrow& \phi 
\end{Eqnarray}
for $U(1)$ gauge group and 
\begin{Eqnarray}
b_{a} &\rightarrow& b_{a} - \partial_{a}\lambda
  +i[\lambda,b_{a}]  \cr 
\varphi &\rightarrow& \varphi+i[\lambda,\varphi] \cr 
\phi &\rightarrow& \phi+i[\lambda,\phi] 
\end{Eqnarray}
for $U(m)$ gauge group. 
For $U(1)$ gauge group case, (\ref{commutativeKillingaction}) 
is simplified :
\begin{Eqnarray}
S&=&
-\frac{1}{4g_{YM}^{2}\rho^{2}}\int d\Omega(
F_{ab}F^{ab}
+8i\frac{\epsilon^{ab}}{\sqrt{g}}F_{ab}\varphi \cr 
&&\hspace{1.5cm}
-2(\partial_{a}\varphi)(\partial^{a}\varphi)
-8\varphi^{2} 
-2\rho^{2}(\partial_{a}\phi)(\partial^{a}\phi) 
+4\rho^{2}\phi^{2} )\cr 
&=&-\frac{1}{4g_{YM}^{2}\rho^{2}}\int 
d^{2}\sigma \sqrt{g}
(F_{ab}F^{ab}
+8i\frac{\epsilon^{ab}}{\sqrt{g}}
F_{ab}\varphi \cr 
&&\hspace{1.5cm}
+2\varphi \triangle \varphi -8\varphi^{2} 
+2\rho^{2}\phi \triangle \phi +4\rho^{2}\phi^{2}) 
\end{Eqnarray}
where 
\begin{equation}
\triangle 
=\frac{1}{\sin\theta}\frac{\partial}{\partial\theta}
(\sin\theta\frac{\partial}{\partial\theta})
+\frac{1}{\sin^{2}\theta}
\frac{\partial^{2}}{\partial\phi^{2}}
\end{equation}
is the Laplacian on a unit sphere.
The scalar field $\phi$ has a negative mass term. 
This negative mass term is originated from the mass term 
in the action (\ref{action}). 
Since $\triangle \sim l(l+1)$ 
respects $SO(3)$ symmetry, 
this action is $SO(3)$ invariant. 

\vspace{0.5cm}
Let us investigate another large $N$ limit with $\alpha$ fixed. 
In this limit, the radius of the fuzzy sphere 
becomes large and 
the noncommutative sphere is expected to become a noncommutative plane. 
By virtue of the $SO(3)$ symmetry of the fuzzy sphere, 
we may consider the theory around the north pole
without loss of generality. 
Around the north pole, 
$\hat{L}_{3}$ can be approximated as $\hat{L}_{3} \sim (N-1)/2$. 
By defining 
$\hat{L}_{i}^{\prime}=\sqrt{\frac{2}{N-1}}\hat{L}_{i}$, 
the commutation 
relation (\ref{su(2)}) becomes 
\begin{equation}
[\hat{L}_{1}^{\prime},\hat{L}_{2}^{\prime}] \sim i. \label{commflat}
\end{equation}
By further defining 
$\hat{x}_{i}^{\prime}= \alpha\hat{L}_{i}^{\prime}$ 
and $\hat{p}_{i}^{\prime}=\alpha^{-1} 
\varepsilon_{ij}\hat{L}_{j}^{\prime}$ 
$(i,j=1,2)$, 
we have 
\begin{equation} 
[\hat{x}_{1}^{\prime},\hat{x}_{2}^{\prime}]=i\alpha^{2}, 
\hspace{0.4cm}
[\hat{p}_{1}^{\prime},\hat{p}_{2}^{\prime}]=i\alpha^{-2}, 
\hspace{0.4cm}
[\hat{x}_{i}^{\prime},\hat{p}_{j}^{\prime}]=i\delta_{ij} 
\end{equation}
and 
\begin{equation} 
\rho^{\prime2}=\hat{x}_{i}^{\prime}\hat{x}_{i}^{\prime}
=\frac{2}{N-1}\rho^{2}=\frac{N+1}{2}\alpha^{2}
\sim \frac{N}{2}\alpha^{2}. 
\end{equation}
In the coordinates of $\hat{x}_{i}^{\prime}$, 
the Plank constant is given by 
\begin{equation}
\frac{4\pi\rho^{\prime 2}}{N} \sim 2\pi\alpha^{2}.  
\end{equation}
We take the following limit to decompactify the sphere 
around the north pole, 
\begin{Eqnarray}
\alpha =\mbox{fixed}, \hspace{0.4cm}  
\rho^{\prime} \rightarrow \infty \hspace{0.2cm} 
(N \rightarrow \infty). \label{flat}
\end{Eqnarray}
$a \star b$ which is defined in (\ref{productmap}) 
becomes the Moyal product
$a \star_{M} b$  
because of (\ref{commflat}) and the Weyl type ordering property 
in (\ref{nonspheri}). 
The following replacement holds in this limit, 
\begin{Eqnarray}
\frac{1}{N}Tr \rightarrow \int \frac{d\Omega}{4\pi} 
= \int \frac{d^{2}x}{4\pi\rho^{2}} 
= \int \frac{d^{2}x^{\prime}}{4\pi\rho^{\prime2}} 
\end{Eqnarray}
and the rotation around the $1(2)$ axis corresponds to 
the translation in the $2(1)$ direction around the north pole, 
\begin{equation}
Ad(\hat{p}_{i}^{\prime}) 
=\frac{1}{\rho^{\prime}}\varepsilon_{ij}Ad(\hat{L}_{j})
\rightarrow \frac{1}{i}\partial_{i}^{\prime}
\hspace{0.2cm} (i=1,2). 
\end{equation}

\noindent 
We can regard $\hat{a}_{3}$ as the scalar field $\hat{\varphi}$ 
around the north pole. 
Since the mass term 
in the action (\ref{action})
drops in this limit, 
the action becomes 
\begin{Eqnarray}
S_{B}&=&-\frac{\alpha^{4}}{2g^{2}}
Tr([\hat{L}_{1}+\rho\hat{a}_{1}, \hat{L}_{2}+\rho\hat{a}_{2}]^{2})
+\frac{\alpha^{4}}{2g^{2}} 
Tr(\rho\hat{\varphi}[\hat{L}_{i}+\rho\hat{a}_{i},
[\hat{L}_{i}+\rho\hat{a}_{i},\rho\hat{\varphi}]]) \cr
&&
+\frac{\alpha^{4}}{2g^{2}} 
Tr(\rho\hat{\phi}[\hat{L}_{i}+\rho\hat{a}_{i},
[\hat{L}_{i}+\rho\hat{a}_{i},\rho\hat{\phi}]]) 
-\frac{\alpha^{4}}{2g^{2}} 
Tr([\rho\hat{\phi},\rho\hat{\varphi}]
[\rho\hat{\phi},\rho\hat{\varphi}]) \cr 
&&
\cr
&=&\frac{\alpha^{4}}{2g^{2}}(\rho^{\prime})^{4}
\{ 
-Tr([-\hat{p}_{2}^{\prime}-\hat{a}_{2}^{\prime}, 
\hat{p}_{1}^{\prime}+\hat{a}_{1}^{\prime}]^{2})
+Tr(\hat{\varphi}^{\prime}
[\hat{p}_{i}^{\prime}+\hat{a}_{i}^{\prime},
[\hat{p}_{i}^{\prime}+\hat{a}_{i}^{\prime},
\hat{\varphi}^{\prime}]]) \cr
&&\hspace{1.5cm} +Tr(\hat{\phi}^{\prime}
[\hat{p}_{i}^{\prime}+\hat{a}_{i}^{\prime},
[\hat{p}_{i}^{\prime}+\hat{a}_{i}^{\prime},\hat{\phi}^{\prime}]]) 
-Tr([\hat{\phi}^{\prime},\hat{\varphi}^{\prime}]
[\hat{\phi}^{\prime},\hat{\varphi}^{\prime}])  
\} 
\end{Eqnarray} 
where we have defined 
$\hat{a}_{i}=\sqrt{\frac{2}{N}}\hat{a}_{j}^{\prime}
\varepsilon_{ji} \hspace{0.1cm}(i,j=1,2)$, 
$\hat{\varphi}=\sqrt{\frac{2}{N}}\hat{\varphi}^{\prime}$ 
and $\hat{\phi}=\sqrt{\frac{2}{N}}\hat{\phi}^{\prime}$.
This action can be mapped to the following 
field theory action, 
\begin{Eqnarray}
S_{B}&=&
-\frac{\alpha^{6}N^{2}}{16\pi g^{2}}
\{ \int d^{2}x^{\prime }F_{12}(x)^{2}
+\int d^{2}x^{\prime}
(D_{i}\varphi^{\prime}(x))(D_{i}\varphi^{\prime}(x)) \cr
&&\hspace{1.5cm} +\int d^{2}x^{\prime}
(D_{i}\phi^{\prime}(x))(D_{i}\phi^{\prime}(x))
+\int d^{2}x^{\prime}
[\phi^{\prime},\varphi^{\prime}]
[\phi^{\prime},\varphi^{\prime} ]
\}_{\star}  , 
\end{Eqnarray}
where $D_{i}=\frac{1}{i}\partial_{i}^{\prime}
+[a_{i}^{\prime}, \cdot]$ and 
$F_{12}=\frac{1}{i} \partial_{1}^{\prime}a_{2}^{\prime}
-\frac{1}{i}\partial_{2}^{\prime}a_{1}^{\prime}
+[a_{1}^{\prime},a_{2}^{\prime}]_{\star}$. 
It is found that 
the Yang-Mills coupling is $g_{YM}^{2}=4\pi g^{2}/N^{2}\alpha^{6}$. 
We thus obtained an action of a gauge theory 
on a noncommutative plane 
by taking a large $N$ limit with 
noncommutativity fixed 
from a gauge theory on a noncommutative sphere. 

The gauge transformation (\ref{matrixgauge}) becomes
\begin{Eqnarray}
a_{i}^{\prime}(x) &\rightarrow &
a_{i}^{\prime}(x)-\partial_{i}^{\prime}\lambda(x)
  +i[\lambda(x),a_{i}^{\prime}(x)]_{\star} \cr  
\varphi^{\prime}(x) &\rightarrow & \varphi^{\prime}(x)
+i[\lambda(x),\varphi^{\prime}(x)]_{\star} \cr 
\phi^{\prime}(x) &\rightarrow & \phi^{\prime}(x)
+i[\lambda(x),\phi^{\prime}(x)]_{\star} .
\end{Eqnarray}


\vspace{0.6cm} 
In our previous paper\cite{IKTW}, 
we consider a matrix model with Chern-Simons term. 
Before finishing this section, 
we remark on a difference between 
CS term and mass term. 
Action which is used in the previous paper is 
\begin{equation}
 S= \frac{1}{g^{2}} Tr ( -\frac{1}{4} [ A_{i} ,A_{j}] 
         [A^{i} ,A^{j} ]  
  +\frac{2}{3}i\alpha \epsilon_{ijk}A^{i}A^{j}A^{k}).
\end{equation}
The difference is given by the following action, 
\begin{equation}
S^{\prime}= \frac{1}{g^{2}} Tr(
\frac{2}{3}i\alpha \epsilon_{ijk}A^{i}A^{j}A^{k}
+\alpha^{2} A_{i}A_{i}) 
\label{differencematrix} 
\end{equation}
and it is mapped to the following field theory action as   
\begin{Eqnarray}
S^{\prime}=\frac{i}{g_{YM}^{2}}
\epsilon_{ijk}\int d\Omega (
(L_{i}a_{j})a_{k}+\frac{\rho}{3}[a_{i},a_{j}]a_{k}
-\frac{i}{2}\epsilon_{ijl}a_{l}a_{k})_{\star} 
+\frac{\pi}{3g_{YM}^{2}}\frac{N^{2}}{\rho^{2}} .
\end{Eqnarray} 
This represents a gauge invariant coupling between 
the gauge field and the scalar field. 
(Gauge invariance is manifest 
in (\ref{differencematrix}).)
In the commutative limit this becomes as follows 
\begin{Eqnarray}
S^{\prime}&=&\frac{1}{g_{YM}^{2}\rho^{2}}
\int d\Omega
(i\epsilon_{ijk}
K_{i}^{a}K_{j}^{b}F_{ab}\frac{x_{k}}{\rho}\varphi 
-\varphi^{2}
)\cr 
&=&\frac{1}{g_{YM}^{2}\rho^{2}}
\int d\Omega
(i\frac{\epsilon^{ab}}{\sqrt{g}}F_{ab}\varphi
-\varphi^{2}).
\end{Eqnarray} 
In considering a field theory on a sphere, 
we may add a term of this type because 
gauge invariance is not lost. 
While a matrix model with Chern-Simons term 
leads to a {\it supersymmetric} gauge theory on a fuzzy sphere 
as discussed in \cite{IKTW}, 
a matrix model with mass term does not. 
 

\section{Noncommutative gauge theory on noncommutative torus}
\hspace{0.4cm}
In this section we consider the second classical solution 
of our model, a noncommutative torus. 
It is described naturally 
in terms of unitary matrices 
(by imposing a condition (\ref{noncommtorus})). 
Since eigenvalues of unitary matrices are distributed 
over $S^{1}$, a compact spacetime appears naturally 
from the unitary matrices. 
In this section, in order to treat a compact spacetime 
in terms of hermitian matrices, we decompose 
a unitary matrix into two hermitian matrices. 
This corresponds to a treatment of a two-dimensional torus 
embedded into ${\bf R^{4}}$.

The algebra on the fuzzy torus is generated by 
two unitary matrices $U$ and $V$ which satisfy 
the following relation, 
\begin{Eqnarray}
UV=e^{i\theta}VU \label{noncommtorus}
\end{Eqnarray}
where $\theta =2\pi/N$. 
It is well known that 
these two traceless unitary matrices 
are given by the following 
$N \times N$ clock and shift matrices, 

\begin{equation}
U= \left( \begin{array}{ccccc}
 1 &   & & & \\
   & \omega & & & \\
   &   &  \omega^{2}& & \\
   &   & & \ddots& \\
   &   & & & \omega^{N-1} \\ 
\end{array} \right)  \hspace{0.4cm} 
V= \left( \begin{array}{ccccc}
 0&1   & & &0 \\
   & 0 &1 & & \\
   &   &  \ddots&\ddots & \\
   &   & & \ddots&1 \\
  1 &   & & & 0 \\ 
\end{array} \right)  
\end{equation}
where $\omega=e^{i\theta} (\omega^{N}=1)$. 
$U$ and $V$ satisfy 
\begin{equation}
U^{N}=V^{N}=1 
\label{unitmatrices}
\end{equation}
and
\begin{Eqnarray}
&&UU^{\dagger}=U^{\dagger}U=1,\hspace{0.4cm}
VV^{\dagger}=V^{\dagger}V=1. \label{unitarycond}
\end{Eqnarray}
To embed two-dimensional torus into {\bf R$^{4}$}, 
we decompose these unitary matrices into hermitian matrices: 

\begin{Eqnarray}
U &\equiv& \hat{y}_{1}+i\hat{y}_{2} \equiv e^{i\hat{u}}  \cr
V &\equiv& \hat{y}_{3}+i\hat{y}_{4} \equiv e^{i\hat{v}} 
\label{defhermitian}
\end{Eqnarray} 
or 
\begin{Eqnarray}
\hat{y}_{1}&=& \frac{1}{2}(U+U^{\dagger}) \hspace{0.4cm}
\hat{y}_{2}= \frac{1}{2i}(U-U^{\dagger}) \cr 
\hat{y}_{3}&=& \frac{1}{2}(V+V^{\dagger}) \hspace{0.4cm}
\hat{y}_{4}= \frac{1}{2i}(V-V^{\dagger}) .
\label{ys}
\end{Eqnarray} 
$\hat{w}\equiv R_{1}\hat{u}$ 
and $\hat{z}\equiv R_{2}\hat{v}$ 
are introduced as {\it noncommutative coordinates} 
on the torus. 
From (\ref{noncommtorus}) and (\ref{defhermitian}), 
$\hat{u}$ and $\hat{v}$  
satisfy the following commutation relation: 

\begin{Eqnarray}
[\hat{u}, \hat{v}]=\frac{2\pi i}{N}=i\theta.
\end{Eqnarray}
This is the same relation as the usual flat case. 
Strictly speaking, 
such $\hat{u}$ and $\hat{v}$ does not exist 
for finite $N$
while $U$ and $V$ exist for finite $N$. 
(\ref{unitarycond}) can be rewritten in terms of $\hat{y}_{i}$ as 
\begin{Eqnarray}
\hat{y}_{1}^{2}+\hat{y}_{2}^{2}=1 , \hspace{0.4cm}
[\hat{y}_{1}, \hat{y}_{2}]=0, \cr
\hat{y}_{3}^{2}+\hat{y}_{4}^{2}=1 , \hspace{0.4cm}
[\hat{y}_{3}, \hat{y}_{4}]=0.
\label{radius1}
\end{Eqnarray}
Commutation relations of $\hat{y}_{i}$'s are
represented by 
\begin{Eqnarray}
[\hat{y}_{1},\hat{y}_{2}]&=&0 \cr
[\hat{y}_{3},\hat{y}_{4}]&=&0 \cr
[\hat{y}_{1},\hat{y}_{4}]&=&
(\cos\theta -1)\hat{y}_{4}\hat{y}_{1}
+i\sin\theta \hat{y}_{3}\hat{y}_{2} \cr
[\hat{y}_{2},\hat{y}_{3}]&=&
(\cos\theta -1)\hat{y}_{3}\hat{y}_{2}
+i\sin\theta \hat{y}_{4}\hat{y}_{1}\cr
[\hat{y}_{1},\hat{y}_{3}]&=&
(\cos\theta -1)\hat{y}_{3}\hat{y}_{1}
-i\sin\theta \hat{y}_{4}\hat{y}_{2}\cr
[\hat{y}_{2},\hat{y}_{4}]&=&
(\cos\theta -1)\hat{y}_{4}\hat{y}_{2}
-i\sin\theta \hat{y}_{3}\hat{y}_{1}. 
\label{commutationtorus}
\end{Eqnarray}
It can be shown that 
$A_{\mu}=\hat{x}_{\mu}\equiv R_{a}\hat{y}_{\mu}$ 
($a=1$ for $\mu=1,2$ and $2$ for $\mu=3,4$) 
satisfy the equation of motion 
(\ref{classicalsolution}) if we set 

\begin{equation}
\lambda^{2}=\frac{\beta^{2}}{2}
\equiv R_{a}^{2}(1-\cos \theta) 
\end{equation}
\noindent where $a=2$ for $\mu=1,2$ and 
$a=1$ for $\mu=3,4$. 
$\beta$ is a quantity which should be compared 
with $\alpha$ in the previous section. 
$R_{a}(a=1,2)$ are radii of two cycles of the torus. 
From (\ref{radius1}), we obtain 
\begin{equation}
\hat{x}_{1}^{2}+\hat{x}_{2}^{2}=R_{1}^{2} , \hspace{0.4cm} 
\hat{x}_{3}^{2}+\hat{x}_{4}^{2}=R_{2}^{2} .
\end{equation}
This classical solution preserves 
$SO(2)\times SO(2)$ symmetry. 
Hereafter we set $R_{1}=R_{2}$ for simplicity. 

In the commutative limit, 
$\hat{x}_{\mu}$ becomes 
the commutative coordinates on the torus: 
\begin{Eqnarray}
&&x_{1}=R\cos u ,\hspace{0.4cm}  
  x_{2}=R\sin u , \cr 
&&x_{3}=R\cos v ,\hspace{0.4cm}
  x_{4}=R\sin v .  
\end{Eqnarray}
The metric tensor of the torus is 
\begin{Eqnarray}
ds^{2}&=&
R^{2}du^{2}+R^{2}dv^{2}\cr 
&\equiv& R^{2}g_{ab}d\sigma^{a}d\sigma^{b}.
\end{Eqnarray}
The Plank constant, 
which represents the area occupied by the 
unit quantum on the torus, is given by  
\begin{equation}
\frac{(2\pi R)^{2}}{N} 
=2\pi R^{2} \theta 
=N\beta^{2}
\label{planktorus}
\end{equation}
In the second equality, we have assumed large $N$ 
such that $\beta=R\theta$. 

We derive a $U(1)$ noncommutative gauge theory on 
a noncommutative torus 
by expanding the matrices around the classical solution: 

\begin{equation}
A_{\mu}=\hat{x}_{\mu}+R^{2}\theta\hat{a}_{\mu}
\label{expandtorus}
\end{equation}
where $\hat{a}_{\mu}$ are propagating fields on the 
torus.
Matrices on the fuzzy torus can expanded as 
\begin{Eqnarray}
\hat{a}
&=&\sum_{n=0}^{N}\sum_{m=0}^{N}
a_{nm}e^{-\frac{\pi i}{N}nm}U^{n}V^{m}  \cr  
&=&\sum_{n=0}^{N}\sum_{m=0}^{N}
a_{nm}e^{in\hat{u}+im\hat{v}} \cr 
&\equiv& \sum_{n=0}^{N}\sum_{m=0}^{N}
a_{nm}\hat{Z}_{nm}.
\label{expandmatrixtorus}
\end{Eqnarray}
$\hat{Z}_{nm}$ forms a complete basis of $N \times N$ 
hermitian matrices. 
$a^{\ast}_{nm}=a_{-n-m}$ comes from 
a hermitian condition on $\hat{a}$. 
The ordering of $\hat{u}$ and $\hat{v}$ 
corresponds to Weyl type ordering. 
Functions on the fuzzy torus is expanded 
in terms of the plane waves
as 
\begin{Eqnarray}
a(u,v)&=&\sum_{n=0}^{N}\sum_{m=0}^{N}
a_{nm}e^{inu}e^{imv} \cr 
&\equiv&\sum_{n=0}^{N}\sum_{m=0}^{N}
a_{nm}Z_{nm}(u,v). 
\label{expandfunctiontorus}
\end{Eqnarray}
where same momentum modes are used as 
(\ref{expandmatrixtorus}). 
From the condition (\ref{unitmatrices}) which expresses 
the periodicity of the torus, 
the momentum $n$ and $m$ is bounded at $N$.  

Now we consider a map from matrices to functions, 
\begin{equation}
\hat{a} \rightarrow a(u,v)
\end{equation}
From (\ref{expandmatrixtorus}) and (\ref{expandfunctiontorus}), 
we obtain an explicit map from matrices to functions as 
\begin{equation}
a(u,v)=\sum_{nm}\frac{1}{N}
Tr(\hat{Z}_{nm}\hat{a})Z_{nm}(u,v).
\end{equation}
The product of matrices is also mapped to 
the so called star product, 
\begin{equation}
a\star b(u,v)=\sum_{nm}\frac{1}{N}
Tr(\hat{Z}_{nm}\hat{a}\hat{b})Z_{nm}(u,v).
\end{equation}  
To guarantee that this map is one to one, 
we take large $N$ limit 
such that typical $n$ is much smaller than $N$.  
(See also \cite{zachos} for the correspondence 
between matrices and functions.) 
$Tr$ over matrices can be mapped to the integration 
over functions as 
\begin{equation}
\frac{1}{N}Tr(\hat{a}) \rightarrow 
\int \frac{dwdz}{(2\pi R)^{2}}a(w,z). 
\end{equation}

We next consider differential operators corresponding 
to the adjoint action of $\hat{x}_{i}$. 
We first investigate the adjoint action of $U$ and $V$. 
It is calculated as
\begin{Eqnarray}
[U,\hat{a}] 
&=&\sum_{nm}a_{nm}e^{-\frac{\pi i}{N}nm}
(1-e^{-im\theta})U^{n+1}V^{m} \cr
&\rightarrow&
e^{iu} (1-e^{-i\theta(\frac{1}{i}\partial v)})a 
\sim e^{iu}\theta\partial_{v} a , \cr 
[U^{\dagger},\hat{a}] 
&=&\sum_{nm}a_{nm}e^{-\frac{\pi i}{N}nm}
(1-e^{im\theta})U^{n-1}V^{m} \cr
&\rightarrow& 
e^{-iu} (1-e^{i\theta(\frac{1}{i}\partial v)})a  
\sim - e^{iu}\theta\partial_{v} a , \cr
[V,\hat{a}] 
&=&-\sum_{nm}a_{nm}e^{-\frac{\pi i}{N}nm}
(1-e^{-in\theta})U^{n}V^{m+1} \cr
&\rightarrow&
-e^{iv} (1-e^{-i\theta(\frac{1}{i}\partial u)})a 
\sim -e^{iv}\theta\partial_{u} a ,\cr 
[V^{\dagger},\hat{a}] 
&=&-\sum_{nm}a_{nm}e^{-\frac{\pi i}{N}nm}
(1-e^{in\theta})U^{n}V^{m-1} \cr
&\rightarrow&
-e^{-iv} (1-e^{i\theta(\frac{1}{i}\partial u)})a
\sim e^{-iv}\theta\partial_{u} a. 
\end{Eqnarray}
We have taken a large $N$ limit such that $n\theta 
\sim n/N \ll 1$. 
After taking this large $N$ limit, compactness of the torus 
can not be observed. 
From these relations, we can obtain 
the adjoint action of $\hat{x}_{i}$: 
\begin{Eqnarray}
[\hat{x}_{1},\hat{a}] &\rightarrow&
\frac{R}{2}(e^{iu} (1-e^{-i\theta(\frac{1}{i}\partial v)})
+e^{-iu} (1-e^{i\theta(\frac{1}{i}\partial v)}))
a(u,v) \cr 
&\sim& -R\theta \sin u
(\frac{1}{i}\frac{\partial}{\partial v})
a(u,v) , \cr 
[\hat{x}_{2},\hat{a}] &\rightarrow&
\frac{R}{2i}(e^{iu} (1-e^{-i\theta(\frac{1}{i}\partial v)})
-e^{-iu} (1-e^{i\theta(\frac{1}{i}\partial v)}))
a(u,v) \cr 
&\sim& R\theta \cos u (\frac{1}{i}\frac{\partial}{\partial v})
a(u,v) , \cr 
[\hat{x}_{3},\hat{a}] &\rightarrow&
-\frac{R}{2}(e^{iv} (1-e^{-i\theta(\frac{1}{i}\partial u)})
+e^{-iv} (1-e^{i\theta(\frac{1}{i}\partial u)}))
a(u,v) \cr 
&\sim& R\theta \sin v (\frac{1}{i}\frac{\partial}{\partial u})
a(u,v) , \cr 
[\hat{x}_{4},\hat{a}] &\rightarrow&
-\frac{R}{2i}(e^{iv} (1-e^{-i\theta(\frac{1}{i}\partial u)})
-e^{-iv} (1-e^{i\theta(\frac{1}{i}\partial u)}))
a(u,v) \cr 
&\sim& -R\theta \cos v (\frac{1}{i}\frac{\partial}{\partial u})
a(u,v) .
\end{Eqnarray}
These are expressed in terms of Killing vectors 
on the torus as 
\begin{Eqnarray}
[\hat{x}_{\mu},\hat{a}]
&\equiv& \beta [\hat{T}_{\mu},\hat{a}] \cr 
&\rightarrow& 
-i\beta K_{\mu}^{a}\partial_{a}
a(u,v) \cr 
&\equiv&\beta T_{\mu}a(u,v) 
\label{derivativetorus}
\end{Eqnarray}
where $\mu=1,2,3,4$ and $a=u,v$. 
The metric tensor on the torus is also 
expressed in terms of the Killing vectors as 
\begin{Eqnarray}
g^{ab}=K_{i}^{a}K_{i}^{b}.
\end{Eqnarray}
The explicit forms of $K_{\mu}^{a}$ are summarized in appendix.
The Laplacian on the torus is given by 
\begin{Eqnarray}
\frac{1}{\theta^{2}R^{4}}(Ad(\hat{x}))^{2}\hat{a} 
&\rightarrow& 
\frac{1}{\theta^{2}R^{4}}
\{R^{2}(1-\cos(\theta \partial_{v})) 
+R^{2}(1-\cos(\theta \partial_{u}))\}a(u,v) \cr 
&\sim& \frac{1}{2R^{2}} 
(\partial_{u}^{2}+\partial_{v}^{2})a(u,v). 
\end{Eqnarray}
Then we expand the action (\ref{action}) 
around the classical solution (\ref{ys})
as in (\ref{expandtorus})
and apply these mapping rule. 
The action becomes
\begin{Eqnarray}
S&=&-\frac{1}{4g_{YM}^{2}}
 \int dwdz F_{\mu\nu} \star F_{\mu\nu} \cr 
&&+ \frac{i}{g_{YM}^{2}R^{4}} \int dwdz
\{R^{2}F_{13}(x_{4}a_{2}+a_{4}x_{2})
-R^{2}F_{14}(x_{3}a_{2}+a_{3}x_{2}) \cr 
&&\hspace{1.5cm}
-R^{2}F_{23}(x_{4}a_{1}+a_{4}x_{1}) 
+R^{2}F_{24}(x_{3}a_{1}+a_{3}x_{1}) \cr 
&&\hspace{1.5cm}
+(T_{2}a_{1}-T_{1}a_{2})(a_{3}x_{4}-a_{4}x_{3}) \cr 
&&\hspace{1.5cm}
+(T_{4}a_{1})x_{2}a_{3}
-(T_{3}a_{1})x_{2}a_{4}  
-(T_{4}a_{2})x_{1}a_{3}
+(T_{3}a_{2})x_{1}a_{4} \cr 
&&\hspace{1.5cm}
+ia_{4}a_{2}x_{4}x_{2}+ia_{3}a_{2}x_{3}x_{2}  
+ia_{4}a_{1}x_{4}x_{1}+ia_{3}a_{1}x_{3}x_{1}  
\}_{\star} \cr
&&+\frac{1}{2g_{YM}^{2}R^{4}} \int dwdz 
\{2(x_{1}x_{3}a_{1}a_{3}+x_{3}x_{1}a_{3}a_{1}
+x_{1}x_{4}a_{1}a_{4}+x_{4}x_{1}a_{4}a_{1} \cr 
&&\hspace{1.5cm}
+x_{2}x_{3}a_{2}a_{3}+x_{3}x_{2}a_{3}a_{2}
+x_{2}x_{4}a_{2}a_{4}+x_{4}x_{2}a_{4}a_{2}) \cr 
&&\hspace{1.5cm}
+x_{3}a_{1}x_{3}a_{1}+x_{1}a_{3}x_{1}a_{3}
+x_{4}a_{1}x_{4}a_{1}+x_{1}a_{4}x_{1}a_{4} \cr 
&&\hspace{1.5cm}
+x_{3}a_{2}x_{3}a_{2}+x_{2}a_{3}x_{2}a_{3}
+x_{4}a_{2}x_{4}a_{2}+x_{2}a_{4}x_{2}a_{4} \cr 
&&\hspace{1.5cm} 
+O(\theta)
\}_{\star}\cr
&&-\frac{1}{2g_{YM}^{2}R^{2}}\int dwdz
a_{\mu}\star a_{\mu}  
-\frac{1}{g_{YM}^{2}}\frac{N^{2}}{2R^{2}} 
\label{actionexpandedtorus}
\end{Eqnarray}
where $O(\theta)$ contains terms which are 
proportional to $\theta$. 
$F_{\mu\nu}$ is the field strength on the torus: 
\begin{Eqnarray}
\hat{F}_{12}&=&\frac{1}{\theta^{2}R^{4}}([A_{1},A_{2}]), \cr 
\hat{F}_{13}&=&\frac{1}{\theta^{2}R^{4}}
 ([A_{1},A_{3}]-(\cos\theta-1)A_{3}A_{1}
+i\sin\theta A_{4}A_{2}), \cr 
\hat{F}_{14}&=&\frac{1}{\theta^{2}R^{4}}
 ([A_{1},A_{4}]-(\cos\theta-1)A_{4}A_{1}
-i\sin\theta A_{3}A_{2}), \cr 
\hat{F}_{23}&=&\frac{1}{\theta^{2}R^{4}}
 ([A_{2},A_{3}]-(\cos\theta-1)A_{3}A_{2}
-i\sin\theta A_{4}A_{1}), \cr 
\hat{F}_{24}&=&\frac{1}{\theta^{2}R^{4}}
 ([A_{2},A_{4}]-(\cos\theta-1)A_{4}A_{2}
+i\sin\theta A_{3}A_{1}), \cr 
\hat{F}_{34}&=&\frac{1}{\theta^{2}R^{4}}([A_{3},A_{4}]). 
\label{fieldstrengthtorus}
\end{Eqnarray}
These are gauge covariant and equal to zero when 
the fluctuating fields vanish. 
The Yang-Mills coupling is defined by
\begin{equation}
g_{YM}^{2}=\frac{(2\pi R)^{2}g^{2}}{\theta^{4}R^{8}N}
=\frac{g^{2}N^{3}}{(2\pi)^{2}R^{6}}. 
\end{equation}
We have so far discussed the $U(1)$ noncommutative gauge theory 
on the fuzzy torus. 
A generalization to $U(m)$ gauge group is realized 
by the same way as the sphere case. 

Here we note the gauge transformation of 
this noncommutative gauge theory. 
The gauge symmetry of the noncommutative gauge theories 
is embedded in the unitary transformation of the matrix model. 
The gauge transformation 
in noncommutative gauge theories 
is obtained from 
the transformation around the fixed background. 
For $U=\exp (i\hat{\lambda})\sim 1+i\hat{\lambda}$ 
in (\ref{gaugetr}), we obtain 
\begin{Eqnarray}
\hat{a}_{\mu}\rightarrow \hat{a}_{\mu} 
-i\frac{1}{R}[\hat{T}_{\mu},\hat{\lambda}]
+i[\hat{\lambda},\hat{a}_{\mu}] .
\end{Eqnarray}
After mapping to functions, we have local gauge symmetry 
\begin{Eqnarray}
a_{\mu}(u,v)\rightarrow a_{\mu}(u,v)
-i\frac{1}{R}T_{\mu}\lambda(u,v)
+i[\lambda(u,v),a_{\mu}(u,v)]_{\star} ,
\label{gaugesymfunctiontorus}
\end{Eqnarray}
where $T_{\mu}$ is a derivative operator which is 
defined in (\ref{derivativetorus}).

In the same way as the fuzzy sphere case, 
two scalar fields which are transverse components 
of two cycles of the fuzzy torus
is given as follows: 
\begin{Eqnarray}
\hat{\phi}_{1}&=&\frac{1}{2\theta R^{2}}
(A_{1}A_{1}+A_{2}A_{2}
-\hat{x}_{1}\hat{x}_{1}-\hat{x}_{2}\hat{x}_{2}) \cr 
\hat{\phi}_{2}&=&\frac{1}{2\theta R^{2}}
(A_{3}A_{3}+A_{4}A_{4}
-\hat{x}_{3}\hat{x}_{3}-\hat{x}_{4}\hat{x}_{4}).
\end{Eqnarray}
These scalar fields transform covariantly 
as adjoint scalars under the gauge transformation: 
\begin{Eqnarray}
\hat{\phi}_{1}&\rightarrow&
\hat{\phi}_{1} +i[\hat{\lambda}, \hat{\phi}_{1}] \cr 
\hat{\phi}_{2}&\rightarrow&
\hat{\phi}_{2} +i[\hat{\lambda}, \hat{\phi}_{2}].
\end{Eqnarray}

%

\vspace{0.3cm}
We then look at a gauge theory on a commutative torus. 
From (\ref{planktorus}), a commutative limit 
is taken by 
\begin{equation} 
R=\mbox{fixed},  \hspace{0.4cm} 
g_{YM}=\mbox{fixed}, \hspace{0.4cm}
N\rightarrow \infty.  
\label{commutativetorus}
\end{equation}
In this limit, $O(\theta)\rightarrow 0$ 
and the star product becomes the ordinary product. 
Four fields $a_{\mu}$ contain a gauge field on 
$T^{2}$ and two scalar fields $\phi_{1}$ and $\phi_{2}$. 
In the commutative theory, we can separate $a_{\mu}$
in terms of Killing vectors and $x_{\mu}$ as  
\begin{Eqnarray}
a_{\mu}(u,v)=\frac{1}{R}K_{\mu}^{a}b_{a}(u,v)
+\frac{x_{\mu}}{R^{2}}\phi_{i}(u,v) .
\label{commutativefieldstorus}
\end{Eqnarray}
where $i=1$ for $\mu=1,2$ and $i=2$ for $\mu=3,4$.
Field strength in (\ref{fieldstrengthtorus}) 
can be rewritten in terms of the gauge field $b_{a}$ 
and the scalar fields $\phi_{1}$ and $\phi_{2}$ as 
\begin{Eqnarray}
F_{12}&=&\frac{1}{R^{2}}D_{v}\phi_{1}, \cr 
F_{13}&=& \frac{1}{R^{4}}( x_{2} x_{4} F_{uv}
  - x_{2} x_{3}(D_{v}\phi_{2})
 - x_{1} x_{4}(D_{u}\phi_{1}) 
  +x_{1} x_{3}[\phi_{1},\phi_{2}]), \cr 
F_{14}&=& \frac{1}{R^{4}}(- x_{2}x_{3} F_{uv}
- x_{2} x_{4}(D_{v}\phi_{2}) 
 + x_{1} x_{3}(D_{u}\phi_{1})
+x_{1} x_{4}[\phi_{1},\phi_{2}]), \cr 
F_{23}&=& \frac{1}{R^{4}}(- x_{1}x_{4} F_{uv}
  + x_{1} x_{3}(D_{v}\phi_{2}) 
 - x_{2} x_{4}(D_{u}\phi_{1})
 +x_{2} x_{3}[\phi_{1},\phi_{2}] ), \cr 
F_{24}&=& \frac{1}{R^{4}}( x_{1}x_{3} F_{uv}
+ x_{1} x_{4}(D_{v}\phi_{2}) 
+ x_{2} x_{3}(D_{u}\phi_{1})
+x_{2} x_{4}[\phi_{1},\phi_{2}]) ,\cr 
F_{34}&=& \frac{1}{R^{2}}D_{u}\phi_{2}, 
\label{commutativefieldstrengthtorus}
\end{Eqnarray} 
where $F_{uv}=\frac{1}{i}\partial_{u}b_{v}
- \frac{1}{i}\partial_{v}b_{u}+[b_{u},b_{v}]$ 
and $D_{a}=\frac{1}{i}\partial_{a}+[b_{a},\cdot]$. 
Then the action becomes 
\begin{Eqnarray}
S&=&-\frac{1}{4g_{YM}^{2}R^{4}}
 tr\int dwdz (
2(D_{a}\phi_{1})(D^{a}\phi_{1})
+2(D_{a}\phi_{2})(D^{a}\phi_{2}) 
+2[\phi_{1},\phi_{2}]^{2} +F_{ab}F^{ab}) \cr
&&+\frac{i}{g_{YM}^{2}}\frac{1}{R^{4}}tr\int 
dwdz \epsilon^{ab}F_{ab}(\phi_{1}+\phi_{2})
+\frac{1}{g_{YM}^{2}}\frac{1}{R^{4}}tr\int 
dwdz(\phi_{1}\phi_{2}  )  \cr 
&=&-\frac{1}{4g_{YM}^{2}R^{4}}
 tr\int dwdz (
2(D_{a}\chi)(D^{a}\chi)+2(D_{a}\psi)(D^{a}\psi)  
+2[\chi,\psi]^{2} +F_{ab}F^{ab}) \cr
&&+\frac{\sqrt{2}i}{g_{YM}^{2}}\frac{1}{R^{4}}tr\int 
dwdz \epsilon^{ab}F_{ab}\chi 
+\frac{1}{2g_{YM}^{2}}\frac{1}{R^{4}}tr\int 
dwdz(\chi^{2}-\psi^{2}) , 
\label{commutativeKillingaction2}
\end{Eqnarray} 
where $\epsilon^{uv}=-\epsilon^{vu}=1$. 
In the second equality, we changed the variable as 
$\chi=(\phi_{1}+\phi_{2})/\sqrt{2}$ and 
$\psi=(\phi_{1}-\phi_{2})/\sqrt{2}$. 
In the commutative case, the gauge transformation 
becomes 
\begin{Eqnarray}
b_{a}(u,v)&\rightarrow&b_{a}(u,v)
-\partial_{a}\lambda(u,v) 
+i[\lambda(u,v),b_{a}(u,v)] \cr 
\phi_{i}(u,v)&\rightarrow& 
\phi_{i}(u,v)+i[\lambda(u,v),\phi_{i}(u,v) ]
\hspace{0.4cm}(i=1,2)
\end{Eqnarray} 
for $U(m)$ case and 
\begin{Eqnarray}
b_{u}(u,v)&\rightarrow&b_{u}(u,v)
-\partial_{u}\lambda(u,v)  \cr 
\phi_{i}(u,v)&\rightarrow& \phi_{i}(u,v) 
\hspace{0.4cm}
\end{Eqnarray} 
for $U(1)$ case. 
(\ref{commutativeKillingaction2}) is further simplified 
for $U(1)$ gauge group, 
\begin{Eqnarray}
S&=&-\frac{1}{4g_{YM}^{2}R^{4}}
 \int d^{2}\sigma\sqrt{g} (F_{ab}F^{ab}
-4\sqrt{2}i\frac{\epsilon^{ab}}{\sqrt{g}}F_{ab}\chi 
+2\chi\triangle \chi -2\chi^{2}
+2\psi\triangle \psi +2\psi^{2}) 
\end{Eqnarray} 
where 
\begin{equation} 
\triangle = \partial_{u}^{2}+\partial_{v}^{2}
\end{equation}
is the Laplacian on a unit torus. 
A tachyonic scalar field $\psi$ appeared. 
This tachyonic mode 
may be related to the instability of 
the matrix model. 
Since $\triangle \sim -(n^{2}+m^{2})$
respects $SO(2)\times SO(2)$ symmetry, 
this action is also invariant under 
$SO(2)\times SO(2)$ symmetry. 

\vspace{0.5cm}

We next consider another large $N$ limit with $\beta$ fixed. 
In this limit, two radii of a torus become large 
and the noncommutative torus is expected to become a noncommutative plane. 
By virtue of the $SO(2)\times SO(2)$ symmetry 
we may consider the theory around 
$\hat{x}_{1}=R$ and $\hat{x}_{3}=-R$.
We rescale as 
$\hat{x}_{i}=\sqrt{\frac{N}{2\pi}}\hat{x}_{i}^{\prime}
=\frac{1}{\sqrt{\theta}}\hat{x}_{i}^{\prime}$. 
The commutation relations (\ref{commutationtorus})
become 
\begin{equation}
[\hat{x}_{2}^{\prime}, \hat{x}_{4}^{\prime}] =i\beta^{2}
\end{equation}
and other commutation relations vanish. 
Defining 
$\hat{x}_{i}^{\prime}=\beta\hat{T}_{i}^{\prime}$, 
$\hat{p}_{2}^{\prime}=-\beta^{-1}\hat{T}_{4}^{\prime}$ 
and $\hat{p}_{4}^{\prime}=\beta^{-1}\hat{T}_{2}^{\prime}$, 
we get 
\begin{equation} 
[\hat{x}_{2}^{\prime},\hat{x}_{4}^{\prime}]=i\beta^{2}, 
\hspace{0.4cm}
[\hat{p}_{2}^{\prime},\hat{p}_{4}^{\prime}]=i\beta^{-2}, 
\hspace{0.4cm}
[\hat{x}_{2}^{\prime},\hat{p}_{2}^{\prime}]=-i,
\hspace{0.4cm}
[\hat{x}_{4}^{\prime},\hat{p}_{4}^{\prime}]=-i.
\end{equation}
In the coordinates of $\hat{x}_{i}^{\prime}$, 
the Plank constant is 
\begin{equation}
\frac{(2\pi R^{\prime})^{2}}{N}=2\pi \beta^{2}, 
\end{equation}
where 
\begin{equation}
R^{\prime 2}=\theta R^{2}=\frac{1}{\theta}\beta^{2}.
\end{equation}
We take the following limit which leads to 
a two-dimensional noncommutative plane, 

\begin{Eqnarray}
\beta =\mbox{fixed}, \hspace{0.4cm}  
R^{\prime} \rightarrow \infty \hspace{0.2cm} 
(N \rightarrow \infty). \label{flattorus}
\end{Eqnarray} 
The following replacements hold in this case,  
\begin{equation}
\frac{1}{N}Tr \rightarrow 
\int \frac{dwdz}{(2\pi R)^{2}}
=\int \frac{dw^{\prime}dz^{\prime}}{(2\pi R^{\prime})^{2}}
=\int \frac{d{x}_{2}^{\prime}d{x}_{4}^{\prime}}
{(2\pi R^{\prime})^{2}}
\end{equation}
and 
\begin{equation} 
Ad(\hat{p}_{i}^{\prime}) \rightarrow 
\frac{1}{i}\frac{\partial}{\partial x_{i}^{\prime}} 
\end{equation} 
where we have used the fact that 
${x}_{2}\sim Ru$ and ${x}_{4}\sim -R(v-\pi), (u,v-\pi \ll 1)$. 

We can regard $\hat{a}_{1}$ and $\hat{a}_{3}$ 
as $\hat{\phi}_{1}$ and $\hat{\phi}_{2}$ respectively. 
Since the mass term drops in this limit, 
the bosonic part of the action (\ref{action}) becomes 
(only $U(1)$ case is treated for simplicity in the 
present discussions. )
\begin{Eqnarray}
S_{B}&=&-\frac{\beta^{4}}{2g^{2}}
Tr([\hat{T}_{2}+R\hat{a}_{2}, 
\hat{T}_{4}+R\hat{a}_{4}]^{2})
+\frac{\beta^{4}}{2g^{2}} 
Tr(R\hat{\phi_{1}}[\hat{T}_{i}+R\hat{a}_{i},
[\hat{T}_{i}+R\hat{a}_{i},R\hat{\phi_{1}}]]) \cr
&& 
+\frac{\beta^{4}}{2g^{2}} 
Tr(R\hat{\phi_{1}}[\hat{T}_{i}+R\hat{a}_{i},
[\hat{T}_{i}+R\hat{a}_{i},R\hat{\phi_{2}}]])
-\frac{\beta^{4}}{2g^{2}} 
Tr([R\hat{\phi_{1}},R\hat{\phi_{2}}]
[R\hat{\phi_{1}},R\hat{\phi_{2}}])  \cr 
&=&\frac{\beta^{8}}{2g^{2}}(\frac{1}{\sqrt{\theta}})^{4}
\{ 
-Tr([\hat{p}_{4}^{\prime}+\hat{a}_{4}^{\prime}, 
\hat{p}_{2}^{\prime}+\hat{a}_{2}^{\prime}]^{2})
+Tr(\hat{\phi_{1}^{\prime}}
[\hat{p}_{i}^{\prime}+\hat{a}_{i}^{\prime},
[\hat{p}_{i}^{\prime}+\hat{a}_{i}^{\prime},
\hat{\phi_{1}^{\prime}}]]) \cr
&& +Tr(\hat{\phi_{2}^{\prime}}
[\hat{p}_{i}^{\prime}+\hat{a}_{i}^{\prime},
[\hat{p}_{i}^{\prime}+\hat{a}_{i}^{\prime},
\hat{\phi_{2}^{\prime}}]]) 
-Tr([\hat{\phi_{1}^{\prime}},\hat{\phi_{2}^{\prime}}]
[\hat{\phi_{1}^{\prime}},\hat{\phi_{2}^{\prime}}])
\} 
\end{Eqnarray}

\noindent 
where we have defined 
$\hat{a}_{2}=\sqrt{\theta}\hat{a}_{4}^{\prime}$,
$\hat{a}_{4}=\sqrt{\theta}\hat{a}_{2}^{\prime}$,
$\hat{\phi_{1}}=\sqrt{\theta}\hat{\phi_{1}^{\prime}}$ 
and $\hat{\phi_{2}}=\sqrt{\theta}
\hat{\phi_{2}^{\prime}}$.
Repeated index $i$ takes $2$ and $4$. 
This action can be mapped to the following 
field theory action, 
\begin{Eqnarray}
S_{B}&=&
-\frac{\beta^{6}N^{2}}{16\pi^{3} g^{2}}
\{ \int d^{2}x^{\prime }F_{24}(x)^{2}
+\int d^{2}x^{\prime}
(D_{i}\phi_{1}^{\prime}(x))
(D_{i}\phi_{1}^{\prime}(x)) \cr
&&\hspace{1.5cm} +\int d^{2}x^{\prime}
(D_{i}\phi_{2}^{\prime}(x))
(D_{i}\phi_{2}^{\prime}(x))
+\int d^{2}x^{\prime}
[\phi_{1}^{\prime},\phi_{2}^{\prime}]
[\phi_{1}^{\prime},\phi_{2}^{\prime} ]
\}_{\star} , 
\end{Eqnarray}
where $D_{i}=\frac{1}{i}\partial_{i}^{\prime}
+[a_{i}^{\prime}, \cdot]$ and 
$F_{24}=\frac{1}{i} \partial_{2}^{\prime}a_{4}^{\prime}
-\frac{1}{i}\partial_{4}^{\prime}a_{2}^{\prime}
+[a_{2}^{\prime},a_{4}^{\prime}]_{\star}$. 
It is found that 
the Yang-Mills coupling is 
$g_{YM}^{2}=4\pi^{3} g^{2}/N^{2}\beta^{6}$. 
We thus obtained an action of a gauge theory 
on a noncommutative plane 
by taking a large $N$ limit with 
fixed noncommutativity 
from a gauge theory on a noncommutative torus. 

The gauge transformation (\ref{gaugesymfunctiontorus}) becomes
\begin{Eqnarray}
&&a_{i}^{\prime}(x) \rightarrow 
a_{i}^{\prime}(x)-\partial_{i}^{\prime}\lambda(x)
  +i[\lambda(x),a_{i}^{\prime}(x)]_{\star} \cr 
&&\phi_{1}^{\prime}(x) \rightarrow \phi_{1}^{\prime}(x)
+i[\lambda(x),\phi_{1}^{\prime}(x)]_{\star} \cr 
&&\phi_{2}^{\prime}(x) \rightarrow \phi_{2}^{\prime}(x) 
+i[\lambda(x),\phi_{2}^{\prime}(x)]_{\star} .
\end{Eqnarray}


\section{Gauge invariant operators on sphere and torus}
\hspace{0.4cm}

It is shown in \cite{IIKK} that a gauge invariant 
operator in noncommutative gauge theories 
can have non-vanishing momentum 
which is proportional to 
the distance between the end-points of the path. 
(See also \cite{AMNS,DasRey,GHI,DhWa}.)
This section is devoted to the analysis of such 
gauge invariant operators on a sphere and a torus. 
We are treating two manifolds which have different 
topology, genus zero and genus one. 
The difference of the topology is discussed.

We first consider a gauge invariant operator on 
a noncommutative sphere. 
From (\ref{coordinatemomentumsphere}) and 
(\ref{su(2)}), 
translation on sphere, that is rotation, 
is realized by the following unitary transformation 
\begin{equation}
e^{i\hat{L}\cdot \Delta\omega}
\hat{a}_{i} (\hat{x}_{i})
e^{-i\hat{L}\cdot \Delta\omega}
=\hat{a}_{i}( 
\hat{x}_{i}+\epsilon_{ijk}\Delta\omega_{j}\hat{x}_{k}) 
\end{equation}
for infinitesimal small value $\Delta\omega$. 
This rotation is obtained from the gauge transformation 
(\ref{gatr}) if we set 
$\hat{\lambda}=\hat{L}_{i}\Delta\omega_{i}$: 
\begin{equation}
a_{i}(x) \rightarrow a_{i}(x) 
+i\Delta\omega_{j}L_{j}a_{i}(x) 
+\frac{1}{\rho\alpha}\epsilon_{ijk}\Delta\omega_{j}x_{k}. 
\end{equation}

Now let us study a gauge invariant operator on the sphere. 
For simplicity we consider a rotation  
in $\hat{x}_{3}=d=$constant plane. 
The generator is $\hat{L}_{3}$. 
A gauge invariant operator is made 
by covariantizing $\exp(i\hat{L}_{3}\omega_{3})$ 
as 
\begin{Eqnarray}
W=\frac{1}{N}Tr \exp(\frac{i}{\alpha}A_{3}\omega_{3}).
\end{Eqnarray} 
We now rewrite this as 
\begin{Eqnarray} 
W&=&\frac{1}{N}Tre^{i(\hat{L}_{3}
   +\rho \hat{a}_{3})\omega_{3}} \cr 
&=& \lim_{n\rightarrow \infty}\frac{1}{N}Tr
e^{i(\hat{L}_{3}+\rho \hat{a}_{3})
\frac{\omega_{3}}{n}} \cdots 
e^{i(\hat{L}_{3}+\rho \hat{a}_{3})
\frac{\omega_{3}}{n}} \cr 
&=& \lim_{n\rightarrow \infty}\frac{1}{N}Tr
e^{i\hat{L}_{3}\frac{\omega_{3}}{n}} 
e^{i\rho \hat{a}_{3}\frac{\omega_{3}}{n}} 
 \cdots 
e^{i\hat{L}_{3}\frac{\omega_{3}}{n}} 
e^{i\rho \hat{a}_{3}\frac{\omega_{3}}{n}} \cr 
&=&\lim_{n\rightarrow \infty}\frac{1}{N}Tr 
\prod _{m=1}^{n}e^{i\frac{\omega_{3}}{n}
\rho\hat{a}_{3}
(\vec{x}+\frac{m}{n}\vec{\omega}\times\vec{x})}
e^{i\frac{\omega}{\alpha}\hat{x}_{3}} \cr 
&\rightarrow&\int\frac{d\Omega}{4\pi}
\exp(i\int_{0}^{\omega_{3}}d\tilde{\omega}_{3}\rho
a_{3}(\vec{x}+\vec{\tilde{\omega}}\times\vec{x}))_{\star}
\star e^{ik_{3} x_{3}} 
\end{Eqnarray}
where $k_{3}=\omega_{3}/\alpha$ and $\vec{\omega}=(0,0,\omega_{3})$. 
Taking account of the fact 
that the value of $x_{3}$ takes 
the integer multiple of $\alpha$ 
(or the half integer multiple of $\alpha$), 
total length of the Wilson line is expressed 
in terms of $k_{3}^{\prime}=\omega_{3}^{\prime}/\alpha
=(\omega_{3}-2\pi)/\alpha$: 
\begin{equation}
l=2\pi\sqrt{\rho^{2}-d^{2}}n_{win}
+\sqrt{\rho^{2}-d^{2}}\alpha k_{3}^{\prime}. 
\end{equation}
$n_{win}$ is an integer and represent the winding number. 
This shows that this operator 
has the momentum which is proportional 
to the distance between two end-points up to winding modes 
and the direction of the momentum is orthogonal 
to the direction of the path. 
These are characteristic features 
of noncommutative gauge theories. 
$l$ becomes $2\pi\sqrt{\rho^{2}-d^{2}}n_{win}$ 
if we take the commutative limit. 
The contour becomes closed and 
$l$ vanishes when $d$ approaches $\rho$. 
This shows that the Wilson loops on a commutative sphere 
is contractable. On the other hand, 
Wilson loops on a torus show a different behavior 
as considered in the next part. 

\vspace{0.5cm}


We next have a discussion of the noncommutative torus. 
Translation on the torus is generated by 
the unitary operators $U$ and $V$: 
\begin{Eqnarray}
U\hat{a}(\hat{x})U^{\dagger}&=&
\hat{a}(\hat{x}_{1},\hat{x}_{2},
\cos\theta\hat{x}_{3}-\sin\theta\hat{x}_{4},
\sin\theta\hat{x}_{1}+\cos\theta\hat{x}_{4}), \cr 
V\hat{a}(\hat{x})V^{\dagger}&=&
\hat{a}(
\cos\theta\hat{x}_{1}+\sin\theta\hat{x}_{2},
-\sin\theta\hat{x}_{1}+\cos\theta\hat{x}_{2},
\hat{x}_{3},\hat{x}_{4},
).
\end{Eqnarray}
This is also expressed in terms of $\hat{u}$ and 
$\hat{v}$, 
\begin{Eqnarray} 
U\hat{a}(\hat{u},\hat{v})U^{\dagger}&=&
\hat{a}(\hat{u},\hat{v}+\theta), \cr 
V\hat{a}(\hat{u},\hat{v})V^{\dagger}&=&
\hat{a}(\hat{u}-\theta,\hat{v}),
\end{Eqnarray}
where $\theta=2\pi/N$. 
These show that 
$U$ and $V$ are translation operators 
in the $v$ and $u$ direction on the torus by angle $\theta$ 
respectively. 
$U^{N}=1$ and $V^{N}=1$ are operators 
which perform full translation around two cycles 
of the torus. 

For simplicity, we treat 
a Wilson line operator 
whose path is extended only with $v$ direction. 
The generalization to an arbitrary path 
is straightforward. 
Translation along $v$ direction is generated by 
the unitary operator $U$. 
A gauge invariant Wilson line operator 
is obtained by covariantizing the 
translation operator $U$ as 
\begin{Eqnarray} 
W&=&\frac{1}{R^{M}N}Tr(A_{1}+iA_{2})^{M} \cr 
&=&\frac{1}{N}Tr (U+R\theta \hat{a}_{+})^{M} 
\end{Eqnarray} 
where $MR\theta$ is length of the path 
($M$ is a integer). 
We rewrite it as follows and 
show that this operator can have momentum 
which is proportional to the 
distance between the two ends of the contour. 
\begin{Eqnarray} 
W&=&
\lim_{n\rightarrow \infty}\frac{1}{N}
Tr (U+R\theta \hat{a}_{+})^{\frac{M}{n}n} \cr 
&=&\lim_{n\rightarrow \infty}\frac{1}{N}
Tr (U+R\theta \hat{a}_{+})^{\frac{M}{n}}
 (U+R\theta \hat{a}_{+})^{\frac{M}{n}} \cdots 
(U+R\theta \hat{a}_{+})^{\frac{M}{n}} \cr 
&=&\lim_{n\rightarrow \infty}\frac{1}{N}
Tr\left(1+\frac{M}{n}U^{\dagger}R\theta 
\hat{a}_{+}\left(\hat{v}+\frac{M}{n}\theta
\right)\right)
\left(1+\frac{M}{n}U^{\dagger}R\theta \hat{a}_{+}
\left(\hat{v}+\frac{M}{n}2\theta
\right)\right) \cr
&&\hspace{1cm}\cdots 
\left(1+\frac{M}{n}U^{\dagger}R\theta \hat{a}_{+}
\left(\hat{v}+\frac{M}{n}n\theta
\right)\right)U^{M} \cr 
&=&\lim_{n\rightarrow \infty}\frac{1}{N}
Tr\prod_{m=1}^{n}e^{\frac{M}{n}U^{\dagger}
R\theta \hat{a}_{+}(\hat{v}+\frac{M}{n}m\theta)}U^{M} \cr
&\rightarrow&
\int \frac{dwdz}{(2\pi R)^{2}}
\exp\left(\int_{0}^{M\theta}Rd\tilde{v}e^{-iu}
 a_{+}(v+\tilde{v})\right)_{\star}
\star e^{ik_{w}w} \label{wilsonlinetorus}
\end{Eqnarray}
where $k_{w}=M/R$ and the following manipulation are done in the 
above calculations, 
\begin{Eqnarray}
&&(U+R\theta \hat{a}_{+})^{\frac{M}{n}}\cr
&=&U^{\frac{M}{n}}
(1+U^{\dagger}R\theta \hat{a}_{+})^{\frac{M}{n}} \cr 
&=&U^{\frac{M}{n}}
\left(1+{\frac{M}{n}}U^{\dagger}
R\theta \hat{a}_{+}\right) \cr 
&=&
\left(1+{\frac{M}{n}}U^{\dagger}R\theta 
\hat{a}_{+}\left(\hat{v}+\frac{M}{n}\theta
\right)\right)
U^{\frac{M}{n}}.
\end{Eqnarray} 
We have assumed that $R\theta=\beta$ is small. 
The momentum which is carried by 
the Wilson line operator is 
not $k_{w}$ but also $k_{w}^{\prime}$ which 
is defined by the following equation,  
\begin{Eqnarray}
l&=&MR\theta = k_{w}R\beta \cr  
 &=&2\pi Rn_{win} +k_{w}^{\prime}R\beta ,  
\end{Eqnarray}
because the space of $w$ 
looks like lattice with lattice spacing $R\theta(=\beta)$. 
$n_{win}$ represents the winding number. 
We find that 
the momentum is proportional to 
the distance between the two end-points 
of the contour up to winding modes 
and  
the direction of the path and 
the momentum is orthogonal. 
Although there seems to appear a extra 
factor $e^{-iu}$ in front of $a_{+}$ 
in (\ref{wilsonlinetorus}), 
it cancels with a factor which comes from 
$a_{+}$ in the commutative limit:  
\begin{Eqnarray}
\int \frac{dwdz}{(2\pi R)^{2}}
\exp\left(\int_{0}^{l/R}d\tilde{v}
\left(
ib_{v}(v+\tilde{v})+\phi_{1}(v+\tilde{v})
\right)
\right) e^{ik_{w}w}. 
\end{Eqnarray}
In the commutative limit the path becomes closed, 
that is $l=2\pi Rn_{win}$. 
This operator is the usual Polyakov loop operator. 
The path is not contractable and is classified by the winding 
number $n_{win}$.


\section{Summary and discussions}
\hspace{0.4cm}

In this paper, we have investigated 
two noncommutative gauge theories on a fuzzy sphere
and a fuzzy torus
in terms of a four-dimensional bosonic matrix model. 
By adding a mass term to the original matrix model, 
this matrix model can describe curved spacetime 
(fuzzy sphere and fuzzy torus). 
By expanding matrices into backgrounds 
and fields propagating on them, 
we obtained noncommutative gauge theories on the backgrounds. 
A characteristic feature of noncommutative gauge theories 
or the matrix model is that 
spacetime and fields are treated on the same footing 
and due to this properties, classical backgrounds 
are deformed by fields on them. 

In the analysis of these noncommutative gauge theories, 
we discussed two large $N$ limits. 
One corresponds to a commutative limits and 
another corresponds to a large radius limits. 
From the matrix model point of view, 
these two gauge theories are equivalent. 
However, some differences appeared 
by taking two large $N$ limits. 
We first studied the commutative limit.
By taking this limit, 
we can obtain  
commutative gauge theories. 
Comparing these two gauge theories, 
the difference of the symmetry appeared. 

The advantage of the compact manifolds is that 
one can construct the solutions in terms of finite size 
matrices 
while a solution which represents a noncommutative plane 
cannot be constructed by finite size matrices.
(From the viewpoint of the field theories, 
$N$ plays the role of the cut off parameter.) 
We have shown that 
gauge theories on a noncommutative plane 
are reproduced from 
gauge theories on 
a noncommutative sphere and a noncommutative torus 
in a large $N$ limit. 

We also discussed gauge invariant operators 
on a sphere and a torus. 
It is well known that 
in noncommutative gauge theories, 
Wilson line operators can be gauge invariant 
by carrying momentum which is proportional to 
the distance between the two ends of the contour.  
In this paper we checked whether 
this fact holds on sphere and torus cases. 
Because these manifolds are compact, 
the Wilson lines on them have winding modes, 
which are topologically different in these two cases. 
One is contractable and another is noncontractable. 
The discussion of the topological feature is 
meaningful only in the commutative limit. 
Since the propagating fields deforms the classical background 
in noncommutative field theories, 
the concept of the topology in noncommutative field theories 
is different from commutative field theories. 
From the viewpoint of the matrix model, 
noncommutative field theories on these manifolds 
are equivalent. By taking the commutative limit, 
the difference of the symmetry 
or the difference of the topology appeared.

One of the future problems is to consider 
noncommutative gauge theories on the other 
noncommutative curved manifolds. 
The extension to higher dimensional manifolds 
is a interesting problem. 
Especially four dimensional case is one of them. 
Since construction of four-dimensional sphere 
in matrix models is already known
in \cite{CST}, it may be possible to 
do the same discussions as this paper. 
Other way of the extension 
to a curved manifold 
is to map onto a complex plane. 
As analyzed in our previous paper \cite{IKTW}, 
we have obtained the normal ordered type basis 
on the fuzzy sphere 
by mapping onto a complex plane. 
After mapping to field theory, 
a product of functions are written 
by the Berezin product\cite{Bere}. 
The merits of using the Berezin product 
is that a star product of a noncommutative manifold 
is given if the K\"{a}hler potential of 
the manifold is given. 
Since the K\"{a}her potential is known for more 
general manifold,  
it may be interesting problem 
to consider more general curved manifolds 
from the matrix model point of view.

\vspace{1cm}
\begin{center}
{\bf Acknowledgments}
\end{center}
\hspace{0.4cm}
We are most grateful to S.Iso for 
helpful discussions and reading 
the manuscript. 
In addition, we would like to 
thank Y.Kitazawa, K.Furuuchi and K.Wakatsuki 
for valuable comments and discussions.

\vspace{0.5cm}

\renewcommand{\theequation}{\Alph{section}.\arabic{equation}}

\appendix
\section{Killing vectors }
\setcounter{equation}{0}
In this appendix, we summarize Killing vectors 
on $S^{2}$ and $T^{2}$. 
\subsection{$S^{2}$}

Killing vectors on $S^{2}$ are given  by 
\begin{Eqnarray}
L_{i}=-iK_{i}^{a}\partial_{a}
\end{Eqnarray}
where 
\begin{equation}
\begin{array}{l l }
K_{1}^{\theta}=-\sin\phi  &
K_{1}^{\phi}=-\cot\theta\cos\phi \\ 
K_{2}^{\theta}=\cos\phi &
K_{2}^{\phi}=-\cot\theta\sin\phi  \\
K_{3}^{\theta}=0  &
K_{3}^{\phi}= 1  . \\
\end{array}
\end{equation}
The metric tensor is written in terms of these vectors as
\begin{Eqnarray}
g^{ab}=K_{i}^{a}K_{i}^{b}.
\end{Eqnarray}
$K_{i}^{a}$ further satisfy the following relation,  

\begin{equation}
\epsilon_{ijk}K_{i}^{\theta}K_{j}^{\phi}
\frac{x_{k}}{\rho}
=\frac{1}{\sin\theta}=\frac{1}{\sqrt{g}}
\end{equation}
where $g=\det g_{ab}$.
\subsection{$T^{2}$}
Killing vectors on $T^{2}$ are given by 

\begin{Eqnarray}
T_{\mu}=-iK_{\mu}^{a}\partial_{a}
\end{Eqnarray}
where 
\begin{equation}
\begin{array}{l l }
K_{1}^{u}= 0 &
K_{1}^{v}=-\sin u  \\ 
K_{2}^{u}=0 &
K_{2}^{v}=\cos u \\
K_{3}^{u}= \sin v  &
K_{3}^{v}=0  \\
K_{4}^{u}= -\cos v & 
K_{4}^{v}=0 . \\
\end{array}
\end{equation}
The metric tensor is written in terms of these vectors as
\begin{Eqnarray}
g^{ab}=K_{i}^{a}K_{i}^{b}.
\end{Eqnarray}




\begin{thebibliography}{99}

\bibitem{pol}
J.Polchinski, 
{\it TASI Lectures on D-Branes}, 
hep-th/9611050.

\bibitem{Taylor} 
W.Taylor, {\it Lectures on D-branes, Gauge Theory and M(atrices)}, 
hep-th/9801182. 

\bibitem{BFSS}
T.Banks, W.Fischler, S.Shenker and L.Susskind, 
{\it M theory as a matrix model: a conjecture}, 
Phys.Rev.D55(1997)5112, hep-th/9610043.

\bibitem{IKKT}
N.Ishibashi, H.Kawai, Y.Kitazawa and A.Tsuchiya, 
{\it A Large $N$ reduced model as superstring}, 
Nucl.Phys.B498(1997)467, hep-th/9612115.

\bibitem{reduced}
T.Eguchi and H.Kawai, 
{\it Reduction of  dynamical degrees of freedom 
in the large N gauge theory}, 
Phys.Rev.Lett.48(1982)1063. 

G.Parisi, 
{\it A simple expression for planar field theories}, 
Phys.Lett.112B(1982)463. 

D.Gross and Y.Kitazawa, 
{\it A quenched momentum prescription for large N theories}, 

Nucl.Phys.B206(1982)440. 

G.Bhanot, U.Heller and H.Neuberger, 
{\it The quenched Eguchi-Kawai model}, 

Phys.Lett113B(1982)47. 

S.Das and S.Wadia, 
{\it Translation invariance and a reduced model 
for summing planar diagrams in QCD}, 
Phys.Lett.117B(1982)228. 

A.Gonzales-Arroya and M.Okawa, 
{\it The twisted Eguchi-Kawai model: A reduced model for large N 
lattice gauge theory}, 
Phys.Rev.D27(1983)2397. 

\bibitem{AIKKT}
H.Aoki, S.Iso, H.Kawai, Y.Kitazawa and T.Tada, 
{\it Space-Time Structures from IIB Matrix Model}, 
Prog.Theor.Phys.99(1998)713, hep-th/9802085.

\bibitem{AIKKTT}
H.Aoki, S.Iso, H.Kawai, Y.Kitazawa, T.Tada and  A.Tsuchiya, 
{\it IIB Matrix Model}, 
Prog.Theor.Phys.Suppl.134(1999)47, hep-th/9908038. 

\bibitem{CDS}
A Connes, M. R. Douglas and A. Schwarz, 
{\it Noncommutative Geometry and Matrix Theory
     : Compactification on Tori}, 
JHEP9802(1998)003, hep-th/9711162. 

\bibitem{SW}
N.Seiberg and E.Witten, 
{\it String theory and Noncommutative Geometry}, 

JHEP9909(1999)032, hep-th/9908056. 

\bibitem{Li}
M.Li, 
{\it Strings from IIB Matrices}, 
Nucl.Phys.B499(1997)149, hep-th/9612222. 

\bibitem{AIIKKT}
H.Aoki, N.Ishibashi, S.Iso, H.Kawai, Y.Kitazawa and T.Tada, 
{\it Noncommutative Yang-Mills in IIB Matrix Model}, 
Nucl.Phys.B565(2000)176, hep-th/9908141. 

\bibitem{IIKK}
N.Ishibashi, S.Iso, H.Kawai and Y.Kitazawa, 
{\it Wilson loops in Noncommutative Yang-Mills} , 
Nucl.Phys.B573(2000)573, hep-th/9910004. 

\bibitem{BM} 
I.Bars, D.Minic, 
{\it Non-Commutative Geometry on a Discrete Periodic Lattice and 
Gauge Theory}, 
hep-th/9910091. 

\bibitem{AMNS}
J.Ambjorn, Y.M.Makeenko, J.Nishimura and R.J.Szabo, 
{\it Finite N Matrix Models of Noncommutative Gauge Theory}, 
JHEP9911(1999)029, hep-th/9911041.   

J.Ambjorn, Y.M.Makeenko, J.Nishimura and  R.J.Szabo, 
{\it Nonperturbative Dynamics of Noncommutative Gauge Theory}, 
Phys.Lett.B480(2000)399, 
hep-th/0002158. 

J.Ambjorn, Y.M.Makeenko, J.Nishimura and R.J.Szabo, 
{\it Lattice Gauge Fields 
and Discrete Noncommutative Yang-Mills Theory}, 
JHEP0005(2000)023, 
hep-th/0004147. 

\bibitem{IKTW} 
S.Iso, Y.Kimura, K.Tanaka and K.Wakatsuki, 
{\it Noncommutative Gauge Theory on Fuzzy Sphere from Matrix Model}, 
hep-th/0101102. 

\bibitem{myers}
R.C.Myers, 
{\it Dielectric-Branes}, 
JHEP9912(1999)022, hep-th/9910053.

\bibitem{Hoppe9702169}
J.Hoppe, 
{\it Some Classical Solutions of Membrane Matrix Model Equations}, 

hep-th/9702169. 

\bibitem{HoppeNicolai}
J.Hoppe and H.Nicolai, 
{\it Relativistic minimal surfaces}, 
Phys.Lett.B196(1987)451. 

\bibitem{Taylor9611042} 
W.Taylor, 
{\it D-brane field theory on compact spaces}, 
Phys.Lett.B394(1997)283. 

\bibitem{bak}
D.Bak and K.Lee, 
{\it Noncommutative Supersymmetric Tubes}, 
hep-th/0103148. 

\bibitem{madore}
J.Madore, 
{\it The Fuzzy Sphere}, 
Class.Quamtum.Grav.9(1992)69.

\bibitem{GKP}
H.Grosse, C.Klimcik and P.Presnajder, 
{\it Towards Finite Quantum Field Theory in Non-Commutative Geometry}, 
Int.J.Theor.Phys.35(1996)231, hep-th/9505175.  

H.Grosse, C.Klimcik and P.Presnajder, 
{\it Field Theory on a Supersymmetric Lattice}, 
Comm.Math.Phys.185(1997)155, hep-th/9507074. 

H.Grosse, C.Klimcik and P.Presnajder, 
{\it Topologically nontrivial 
field configurations in noncommutative geometry}, 
Comm.Math.Phys.178(1996)507, hep-th/9510083. 

\bibitem{WW}
U.Carow-Watamura and S.Watamura, 
{\it Noncommutative Geometry and Gauge Theory on Fuzzy Sphere}, 
Comm.Math.Phys.212(2000)395, hep-th/9801195.

U.Carow-Watamura and S.Watamura, 
{\it Differential Calculus on Fuzzy Sphere and Scalar Field}, 
Int.J.Mod.Phys.A13(1998)3235, q-alg/9710034.

U.Carow-Watamura and S.Watamura, 
{\it Chirality and Dirac Operator on Noncommutative Sphere}, 
Comm.Math.Phys.183(1997)365, hep-th/9605003.

\bibitem{Klimcik}
C.Klimcik, 
{\it Gauge theories on the noncommutative sphere}, 

Comm.Math.Phys.199(1998)257, hep-th/9710153.

\bibitem{GM}
H.Grosse and J.Madore, 
{\it A noncommutative version of the Schwinger model}, 

Phys.Lett.B283(1992)218.

\bibitem{GP}
H.Grosse and P.Presnajder, 
{\it The Dirac Operators on the Fuzzy Sphere}, 

Lett.Math.Phys.33(1995)171. 

\bibitem{bala}
S.Baez, A.P.Balachandran, S.Vaidya and B.Ydri, 
{\it Monopoles and Solitons in Fuzzy Physics}, 
Comm.Math.Phys.208(2000)787, 
hep-th/9811169. 

A.P.Balachandran and S.Vaidya, 
{\it Instantons and Chiral Anomaly in Fuzzy Physics}, 

hep-th/9910129. 

A.P.Balachandran, X.Martin and D.O'Connor, 
{\it Fuzzy Actions and their Continuum Limits}, 
hep-th/0007030. 

\bibitem{KabatTay}
D.Kabat and W.Taylor,  
{\it Spherical membranes in Matrix theory}, 

Adv.Theor.Math.Phys.2(1998)181, hep-th/9711078. 

\bibitem{Rey}
S-J.Rey, 
{\it Gravitating M(atrix) Q-Balls}, 
hep-th/9711081. 

\bibitem{Hoppe}
J.Hoppe, 
{\it Quantum Theory of A Massless Relativistic Surface 
and A Two-Dimensional Bound State Problem}, 
MIT Ph.D.Thesis,1982. 

J.Hoppe, 
{\it Diffeomorphism Groups, Quantization, And $SU(\infty)$}, 

Int.J.Mod.Phys.A4(1989)5235. 

\bibitem{zachos}
D.Fairlie, P.Fletcher and C.Zachos, 
{Trigonometric structure constants 
for new infinite-dimensional algebras}, 
Phys.Lett.B218(1989)203.

D.Fairlie and C.Zachos, 
{\it Infinite-dimensional algebras, sine brackets, and SU($\infty$)}, 

Phys.Lett.B224(1989)101.

D.Fairlie, P.Fletcher and C.Zachos, 
{\it Infinite-dimensional algebras and a trigonometric basis 
for the classical Lie algebras} 
J.Math.Phys.31(1990)1088.

\bibitem{DasRey}
S.Das, S-J.Rey, 
{\it Open Wilson Lines in Noncommutative Gauge Theory 
and Tomography of Holographic Dual Supergravity}, 
Nucl.Phys.B590(2000)453, 
hep-th/0008042. 


\bibitem{GHI}
D.J.Gross, A.Hashimoto and N.Itzhaki, 
{\it Observables of Non-Commutative Gauge Theories}, 
hep-th/0008075. 

\bibitem{DhWa}
A.Dhar and S.R.Wadia, 
{\it A Note on Gauge Invariant Operators 
in Noncommutative Gauge Theories and the Matrix Model}, 
Phys.Lett.B495(2000)413, 
hep-th/0008144. 

\bibitem{CST}
J.Castelino, S.Lee and W.Taylor, 
{\it Longitudinal 5-branes as 4-spheres in Matrix theory}, 
Nucl.Phys.B526(1998)334, 
hep-th/9712105. 

\bibitem{Bere}
F.A.Berezin, 
{\it General Concept of Quantization}, 
Comm.Math.Phys.40(1975)153.




\end{thebibliography}
\end{document}